\documentclass[preprint,12pt]{elsarticle}
\usepackage{amssymb}
 \usepackage{amsthm}
 \usepackage{amsmath}
  \usepackage{booktabs} 
 \usepackage{color}
 \usepackage{xcolor}
 \usepackage{tikz}
 \usetikzlibrary{fit,positioning}
 \usepackage{graphics}
 \usetikzlibrary{arrows}
  \usepackage{algorithmic}
 \usepackage{algorithm}
 \usepackage{subcaption}
 \usepackage{lineno}
\usepackage{float} 
\usepackage{hyperref}
\usepackage{tikz}
\usetikzlibrary{arrows,positioning,shadows}
\usepackage{standalone}
\journal{Knowledge Based Systems}
\DeclareMathOperator{\cut}{\texttt{cut}}
\DeclareMathOperator{\links}{\texttt{links}}

\begin{document}

\begin{frontmatter}

\title{IEDC: An Integrated Approach for Overlapping and Non-overlapping Community Detection}

 \author[rv1]{Mahdi Hajiabadi}
 \author[rv1]{Hadi Zare \corref{cor1}}
 \author[rv1]{Hossein Bobarshad}
 \cortext[cor1]{Corresponding author}
 \address[rv1]{University of Tehran, Iran}
\begin{abstract}
Community detection is a task of fundamental importance in social network analysis that can be used in a variety of  knowledge-based domains. 
While there exist many works on community detection based on  connectivity structures, they suffer from either considering the overlapping or non-overlapping communities. In this work, we propose a novel approach for  general community detection through an integrated framework to extract the overlapping and non-overlapping community structures without assuming prior structural connectivity on networks. Our general framework is based on a primary node based criterion which consists of  the internal association degree along with the external association degree. The evaluation of the proposed method is investigated through the extensive simulation experiments and several benchmark real network datasets. 
The experimental results show that the proposed method outperforms the earlier state-of-the-art algorithms based on the well-known evaluation criteria. 
\end{abstract}

\end{frontmatter} 
\section{Introduction}
Identifying  communities is one of the most fundamental tasks in the network science. The detection of  community structures  has allowed  us to study and discover  the latent underlying mechanism  behind the relationships of the entities of networks. Due to the importance of the communities, there has been a wide range of different applications of community detection including  cultural scene detection \cite{hamdaqa_cultural_2014},  reality epidemic spreading modeling based on community structures \cite{ren_epidemic_2014}, designing network protocols in delay tolerant networks \cite{lu_algorithms_2015}, the pain circulation analysis in tweeter and its effect on the pain therapy  \cite{ tighe_painful_2015}, detecting hierarchical structure of communities for interactive recommendation \cite{chen_detecting_2015}, the impact of physician communities in patient-centric networks on health-care issues \cite{uddin_exploring_2016},  and  revealing  cancer drivers based on the detection of gene communities \cite{cantini_detection_2015}.\par
The community detection  can be considered as an unsupervised learning problem. While the detection of communities is very similar to the unsupervised clustering situation, the network representation of the data resulted in the undesirable behavior of the community detection methods based on typical clustering  techniques such as hierarchical algorithms \cite{girvan2002community}.
On the one hand, applying the tools of network science enhanced the accuracy of the community detection algorithms. 
On the other hand,  complexity and ill-posed essence of the community detection problem have been caused many efforts to get the better solution on a wide variety of applications \cite{fortunato2010community}. There exist different categories of the community detection algorithms.  The most well-known are based on the overlapping and non-overlapping structures of the communities \cite{yang_defining_2015}.  Due to the non-overlapping  assumption of the community structures  in some applications and sound theoretical  roots  in graph theory known as graph coloring problem \cite{fortunato2010community}, it has been attracted many works in this category such as  \cite{newman2006modularity, radicchi_defining_2004, blondel2008fast, meo_generalized_2011, zardi_on2_2013}.\par
Generally, individuals in networks tend to belong  more than one community such as the connections of a person in a social network to several groups of classmates, colleagues and friends or the relationships of  a researcher to variety of fields and collaborators \cite{xie_overlapping_2013, ren_simple_2009}. Based on the overlapping behavior of the community structures, many works are developed  to extract the hidden communities via a variety of different approaches. The examples in this category include detection of connected overlapping communities based on searching the adjacent cliques \cite{lancichinetti_detecting_2009},  edge partitioning techniques  \cite{ahn_link_2010}, label propagation algorithms \cite{liu_discovering_2016}, overlapping community detection in two-mode networks \cite{cui_uncovering_2014, cui_detecting_2016, wang_asymmetric_2016}, topic oriented community detection via a link analysis  approach \cite{zhao_topic_2012},  node location analysis to detect overlapping communities \cite{zhi-xiao_overlapping_2016}, overlapping local neighborhood ratio \cite{eustace_community_2015, eustace_overlapping_2015}, maximal subgraphs \cite{cui_detecting_2014, cui_detecting_2014-1}, detecting core nodes among the communities \cite{wang_detecting_2013},  and model based clustering ideas \cite{handcock_model-based_2007, sun_incorder_2014}. 
Due to the unsupervised essence of the problem,  heuristic and meta-heuristic approaches are investigated to uncover the community patterns through various heuristic fitness functions and nature-inspired algorithms \cite{liu_multiobjective_2014, bijari_memory-enriched_2016}. \par
The earlier works on community detection are suffered from the sensitivity of the initialization of the algorithms, weak accuracy, and leading to many number of communities with singletons. Moreover, most of the earlier approaches have tried to discover the communities either based on overlapping assumption of the community structures or via non-overlapping ones. While these methods resulted in good performance on a variety of dataset adopted to the predefined assumption of the overlapping, they have failed to devise a general framework. Indeed, the communities can be formed by overlapping and non-overlapping structural elements.   In this paper, we propose a novel generalized approach for community detection via a primary association node based criterion based on the combination of  the internal and external association degree. The proposed approach not only considers the overlapping structures of the network communities through an external association degree, but it also applies the non-overlapping patterns of the community structures via  computing the internal association degree for each members of the network. The  experimental results based on extensive stochastic simulation of the networks and several benchmark network datasets verify the superiority of the proposed approach and its strength for general community detection with no specified assumption about the network structure.\par
In the following,  related works on overlapping community detection and a basic idea of the proposed approach are presented in Section \ref{Sec2}. The preliminary concepts and relevant definitions to our method are described in Section \ref{Sec3}.  Section  \ref{Sec4}  introduces the  proposed method for community detection. We illustrate the experimental setting based on the stochastic generated networks and present the evaluation of the proposed algorithm based on the extensive simulated networks  and the real benchmark networks in Section \ref{Sec5}. Section \ref{Sec6} concludes the paper with some suggestions for future works on this ongoing research field. 
\section{Related works and basic idea}
\label{Sec2}
\subsection{Related works}
Many researches have been done on various aspects of the community detection topic. The main contributions in the community detection algorithms are initiated based on the connectivity structures of the networks,  overlapping and non-overlapping ones \cite{yang_defining_2015}. The typical non-overlapping approaches to community detection had been aimed at dividing the network into subnetworks with densely connected internally and weakly connected externally to the others \cite{fortunato2010community}, although some recent overlapping community detection works are aimed to partition the networks to subnetworks with more dense external connections and less dense internal ones \cite{yang2012community}. While the extension of the  non-overlapping methods to detect the overlapping structures would be hardly feasible for most of the algorithms ( and vice versa), the primary aim of the two main paradigms is to extract the hidden community structures consisting up the overlapping and non-overlapping connectivity patterns.
Moreover, the  algorithms in the non-overlapping category fail to discover the overall community structures \cite{newman2006modularity, girvan2002community, fortunato2010community} due to algorithmic restraint and the same problems occurs for the category of overlapping algorithms \cite{ lancichinetti_detecting_2009, airoldi_mixed_2008, yang2012community, whang_overlapping_2016}. The main problems for the aforementioned methods are the high sensitivity to the connectivity structures, leading to many communities with singletons, and  scalability problems for high-dimensional networks. \par
There exist some general community detection algorithms  including the overlapping and hierarchical  detection of community structures\cite{lancichinetti_detecting_2009}, the fuzzy identification of the community structures \cite{sun_identification_2011}, optimizing several objective functions to find communities based on a general search strategy  \cite{sobolevsky_general_2014-3},   interaction-based edge clustering to detect overlapping and hierarchical communities \cite{kim_detecting_2015}. Also, there exist a variety of other works for data representation and local hidden structure detection techniques including structured subspace learning \cite{li_robust_2015},  structure learning based on probabilistic feature selection approach \cite{zare_relevant_2016}, subspace decomposition of the topological relationship among web pages \cite{eustace_approximating_2014},  the overlapping community detection based on maximal clique approach \cite{li_uncovering_2014}  and clustering-guided sparse structural learning  \cite{li_clustering-guided_2014} that are related to our idea on demonstration an integrated framework to detect hidden community structures in network datasets. In the following, we describe the elements of a novel generative probabilistic approach to detect general community structures that could be a positive impact on the performance of the method when compared to the non-generative methods including \cite{lancichinetti_detecting_2009, sun_identification_2011, sobolevsky_general_2014-3} and it has less computational complexity than the  \cite{sobolevsky_general_2014-3, kim_detecting_2015}.   
\subsection{The basic idea of the proposed approach}   
The main question is how one can design a general framework to extract the community structures without either solely considering the overlapping connectivity patterns and or only the non-overlapping ones. We propose a unified  approach based on the principle of  \emph{``division of each community into  the  non-overlap and overlap parts''} providing  a more robust --less sensitive-- results as compared with the earlier works.  Our idea is based on the formation of  network communities by considering both of the overlapping and non-overlapping connectivity constituents. The proposed approach considers a generative stochastic modeling for the overlapping extraction of the communities and a feature extraction approach for internal community structures. Our aim is to construct a general community detection approach in a probabilistic model based framework to detect communities on overlapping and non-overlapping connectivity structures with better accuracy and acceptable computational complexity than the similar ad-hoc overlapping or non-overlapping procedures.\par
We design an integrated framework through an association node based approach which comprises two elements for considering the whole structure of the connectivity patterns of the network.  Indeed,  two parameters are regarded as the non-overlap and overlap parts of community memberships for each nodes. The first parameter ``\emph{IA}'' exploits the local neighbors connectivity behavior of a given node to measure the internal association degree of that node for every community \cite{tang_scalable_2012}. Although, internal association is suitable to detect the non-overlapping part of community,  it acts poorly for covering the overlapping regions of communities. 
The second parameter in the proposed approach measures the amount of interactions between communities for each node of a network via a generative stochastic block-models, ``\emph{EA}'',  approach\cite{airoldi_mixed_2008}. 
So, the probability of one node  belongs to the specific community depends on the neighbors of it and the interaction between communities which is calculated by the generative stochastic block-models.
One node belongs to the specific community based on the following community association degree's,  
\begin{itemize}
\item[(i)] Internal association to capture the non-overlapping part of communities
\item[(ii)] External association to uncover the overlapping part of communities
\end{itemize}
We illustrate the proposed approach in the following sections.  
\section{Elements of the proposed approach}
\label{Sec3}
Let us consider the general network with the graph representation $G=(V, E)$ where  the set of nodes and edges are denoted by $V$ and $E$. The community detection problem can be stated as discovering groups of nodes like $c_i$ that can be written as $C = c_1\cup c_2\cup \cdots \cup c_k$.
By introducing the essential concepts, we give now the relevant definitions to our proposed approach. 
\subsection{Internal Association for non-overlapping regions}
Internal association denotes the affinity level of each node to the specific community  which is defined as, 
\begin{equation}
IA_{v_i} (c_i) = \frac{\sum_{v_j\in N(v_i)} P(v_j|c_i) }{|N(v_i)|}
\label{eq:IA}
\end{equation}
where $N(v_i)$ is the neighbors of node $v_i$,  $P(v_j|c_i)$ is the community propagation probability that node $v_j$ belongs to  community $c_i$ which derived from the edge clustering approach \cite{tang_scalable_2012} and $|N(v_i)|$ equals to the number of neighbors of node $v_i$. Intuitively, the relation  \eqref{eq:IA} states that a node's level of affinity through a given community depends on the affinity levels of its neighbors with that  community.
\subsection{External association for overlapping regions}
Block models are the  significant tools in statistical theory and social network analysis  to find groups of nodes with common properties \cite{airoldi_mixed_2008}. According to this model, let suppose to have a network with $n$ nodes and $m$ edges, the aim is to find the labels of each nodes, $v_i\in \{c_i\}_{1}^{k}$ and the interaction matrix $\beta_{k\times k}$ to compute the community level interactions. Intuitively, the $\beta_{c_1,c_2}$ parameter is the number of links between communities $c_1$ and $c_2$. According to block-models, the overlapping level of two communities depends on the amount of interactions between them such that more interactions lead to more overlapping communities behavior. 
\begin{figure}[H]
	\centering
	\begin{minipage}{0.45 \textwidth}
		\centering
		\includegraphics[width=0.9\linewidth]{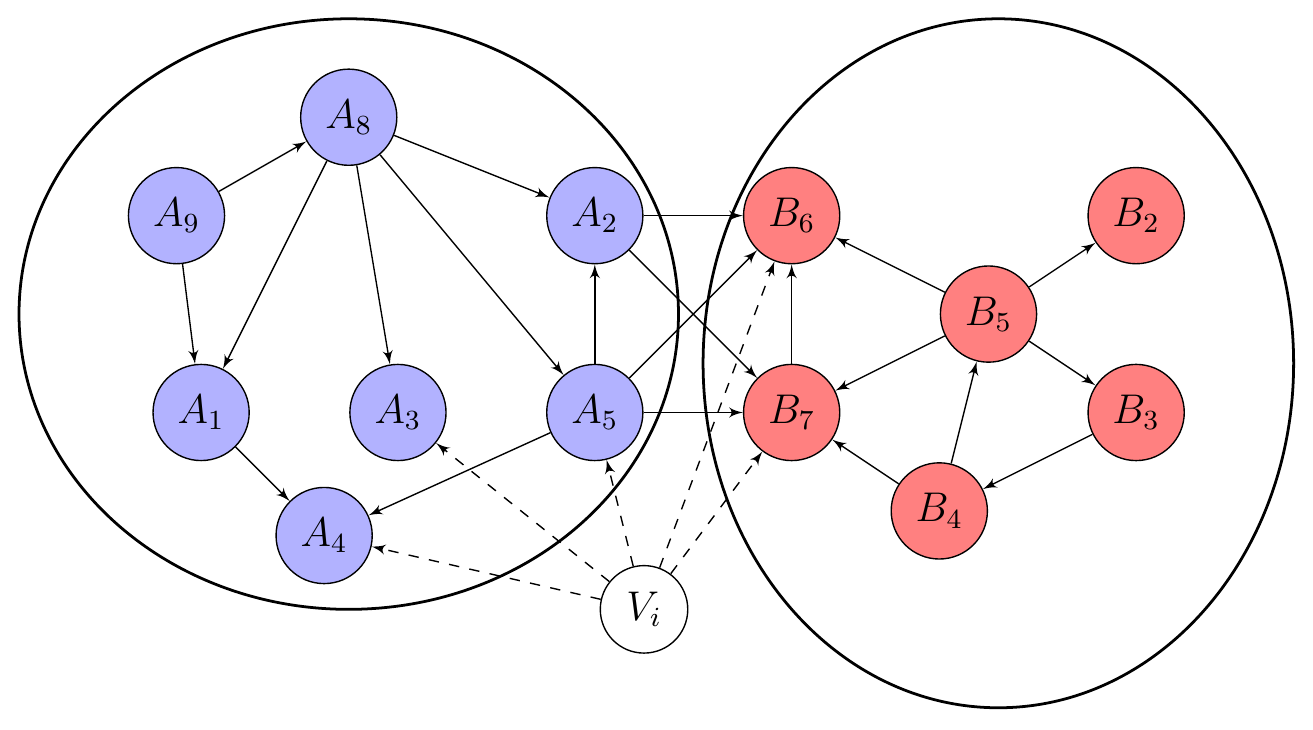}
		\subcaption{}
		\label{subfig:BlockHigh}
	\end{minipage}%
	\begin{minipage}{0.45\textwidth}
		\centering
		\includegraphics[width=0.9\linewidth]{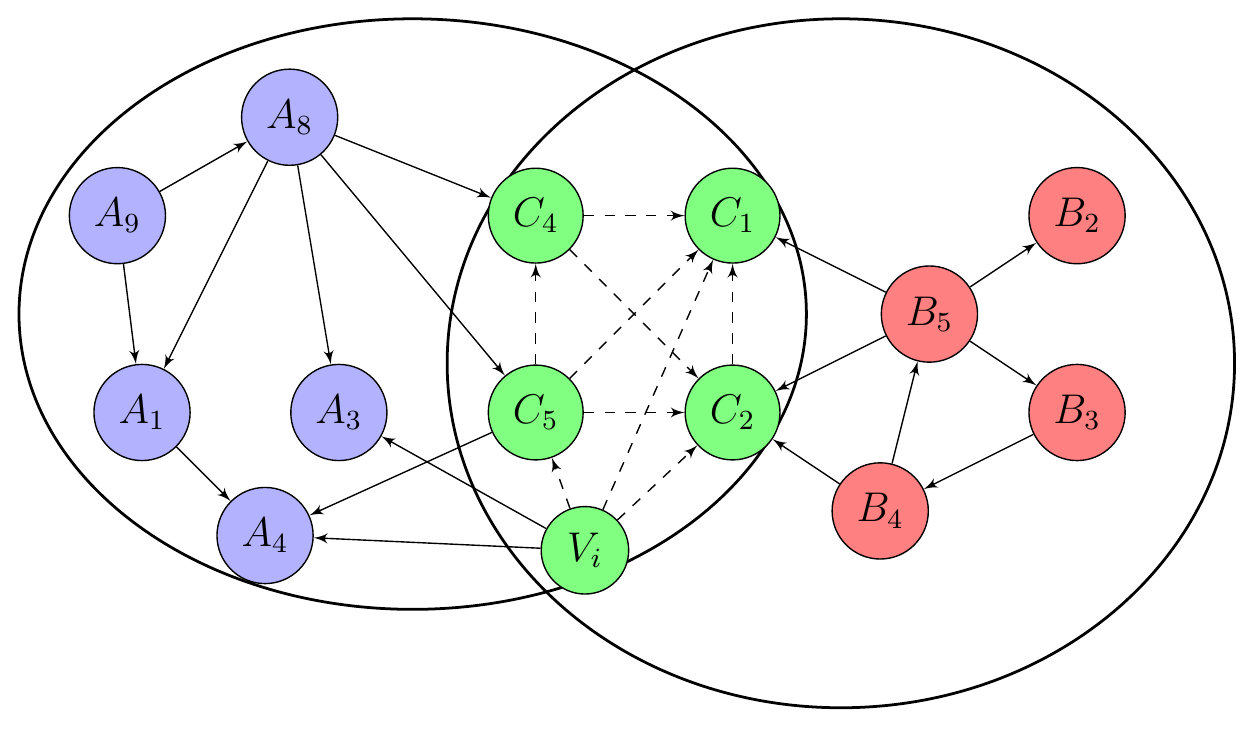}
		\subcaption{}
		\label{subfig:overlapcom}
	\end{minipage}
	\caption{Handling external association degree of communities where Sub-Figure \ref{subfig:BlockHigh} shows two non-overlapping communities with high interactions will become to two overlapping communities in Sub-Figure \ref{subfig:overlapcom}. (Green nodes are overlapping nodes.)}
	\label{Fig.7}
\end{figure}
Figure \ref{Fig.7} represents how higher interactions between communities results in denser overlapping areas. The block-models can be efficiently devised to detect the external association degree of each node to the other communities. The external association degree of each node is computed based on the following two-stage process, 
\begin{itemize}
\item[(i)] Estimate the interaction matrix between any pair of communities 
\item[(ii)] Compute the external association degree of the given node based on the community propagation probability of its neighbors and the interaction matrix.
\end{itemize}
	\begin{figure}[h!]%
		\centering{
		\includegraphics[width=0.9\linewidth]{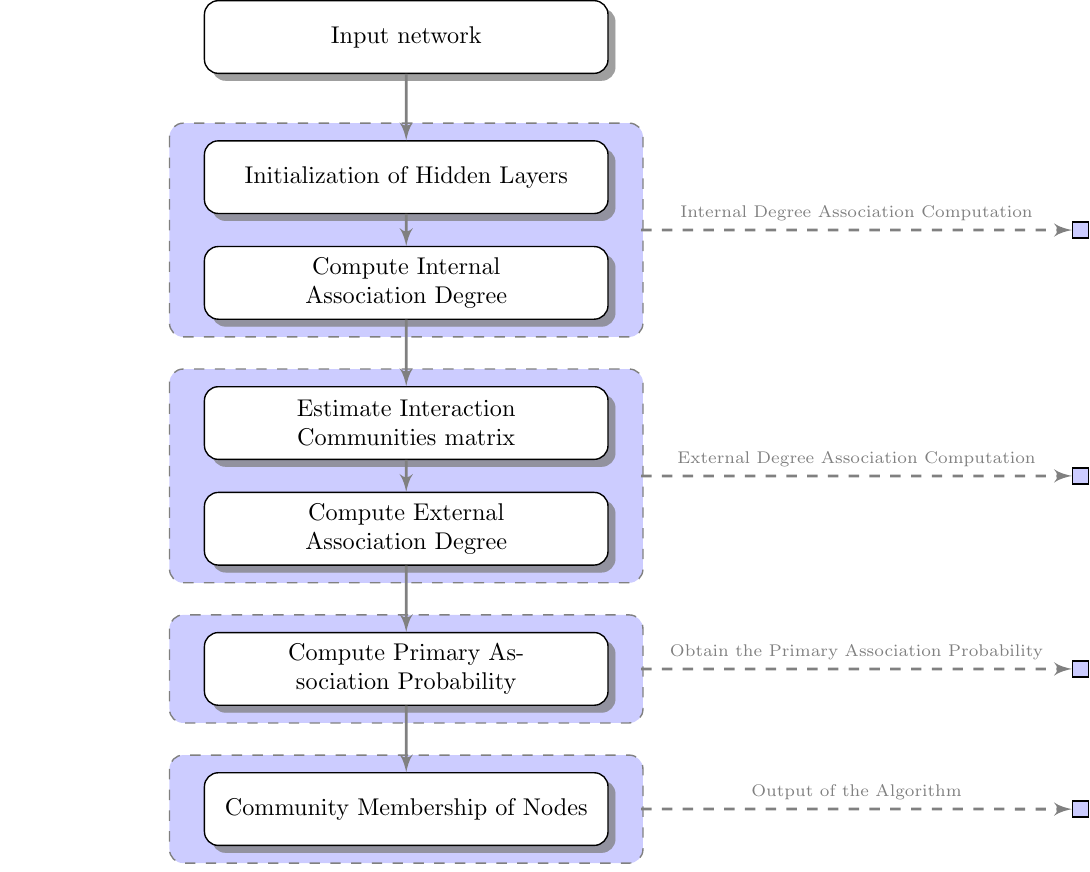}	}
		\caption{The flow-graph of the proposed IEDC algorithm}
		\label{figplan}
	\end{figure}
To detect the communities based on our proposed method, the internal association degree for each node  is computed aligned with the statistical estimation of the external association degree. The flow-graph of proposed algorithm  is shown in Figure \ref{figplan} based on the integration of  ``internal and external association to detect communities'', called as \emph{IEDC}. 
\section{Proposed method for overlapping community detection}
\label{Sec4}
The networks consist up a set of interacting communities, which is the primary assumption behind the summarization of them through the community structures.  Each node of a network would have two-fold association with communities, the first is related to the internal relationship with the other nodes through a community and the second is related to the external relationship with the  nodes from other communities.  While, there exist a few works on the detection of the  general community structures \cite{lancichinetti_detecting_2009, sobolevsky_general_2014-3, kim_detecting_2015}, most of the algorithms have been applied  the internal or external  association of the nodes  to uncover the hidden communities in networks that is known as non-overlapping or overlapping techniques in the literature. We propose an integrated probabilistic method  to detect the community structures via exploiting both of the internal and external association degree for each node of the network.
\subsection{Integrated probabilistic approach}
We proposed the following  probabilistic membership criterion for each node $v_i$ to belong community $c_i$,
\begin{align}
P(v_i\in c_i) &= P(v_i\in c_i| N(v_i)\in c_i) +  P(v_i\in c_i|N(v_i)\in c_{-i}) \label{eq:main1}\\
                    &= p_1(c_i).IA_{c_i}(v_i) + p_2(c_i).EA_{c_i}(v_i) \label{eq:main2}
\end{align}
where $c_{-i}=\bigcup_{j=1}^{k} c_j \setminus c_i$ which the neighbors of $v_i$ belong to them and $p_1(c_i)$ and $p_2(c_i)$ are the importance level of each associated component with it. 
\begin{figure}[h]
	\centering
	\includegraphics[scale = 0.7]{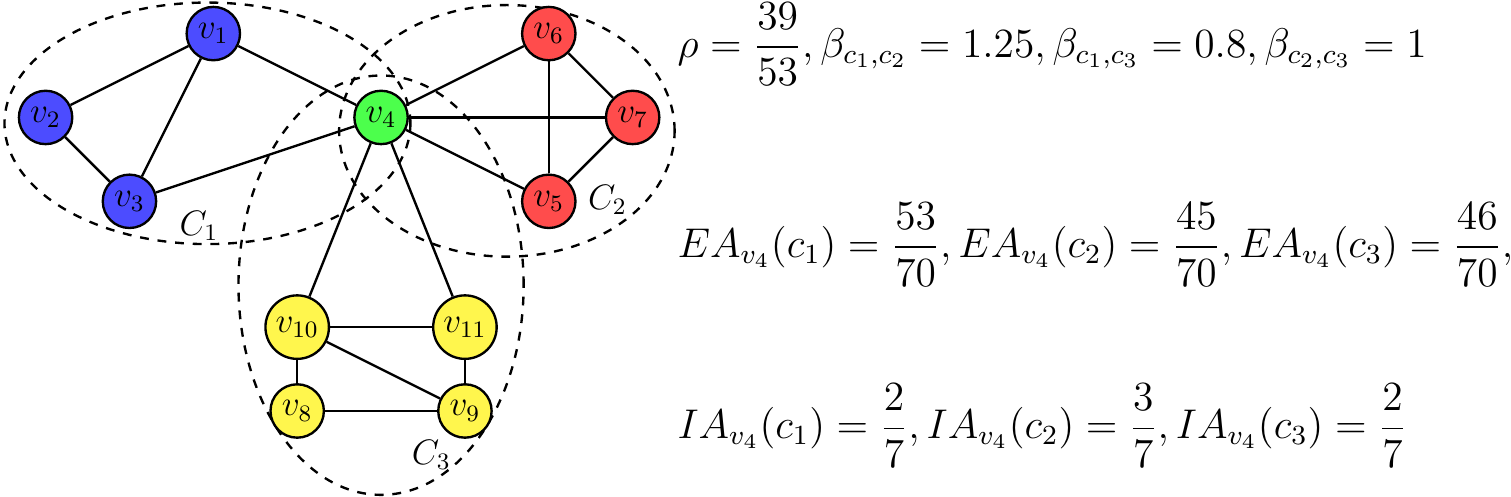}
	\caption{A network containing three communities. Green node exists in three communities $c_1$, $c_2$ , $c_3$. Internal association and external association of node $v_4$ is computed.}
	\label{fig:BlcokExample}
\end{figure}
The first component in \eqref{eq:main1} computes the internal association of node $v_i$ in community $c_i$. The second one reveals the impact of external relationship of communities of  $N(v_i)$ with the current community where higher degree of interaction yields to more overlapping effect.  We apply a generative probabilistic approach to derive the external association of each node as the second component in the main criterion \eqref{eq:main2} based on \cite{airoldi_mixed_2008}. The external association of node $v_i$ in community $c_i$, $EA_{c_i}(v_i)$ is defined as,
\begin{equation}
\label{eq:Overlapping}
EA_{c_i}(v_i) = \frac{\sum_{v_j \in N(v_i)}\sum_{c_j\in C} P(v_j|c_j) \times \beta_{(c_i,c_j)}}{|N(v_i)|}
\end{equation}
where $\beta_{(c_i,c_j)}$ is the interaction matrix between communities. The  interaction matrix  between any pair of communities  can be statistically approximated via the maximum likelihood approach as, 
\begin{equation}
\label{eq:BlockMle}
\beta_{(c_i,c_j)} = \frac{\sum_{(v_i, v_j) \in E} max\Big(P(v_i|c_i) \times P(v_j|c_j),P(v_i|c_j) \times P(v_j|c_i)\Big)}{(1-\rho).\sum_{v_i, v_j}max\Big(P(v_i|c_i) \times P(v_j|c_j),P(v_i|c_j) \times P(v_j|c_i)\Big)} 
\end{equation} 
where $\rho$ is known as the sparsity regularization factor and is computed based on Equation \ref{eq:sparsity} \cite{airoldi_mixed_2008},

\begin{equation}
\label{eq:sparsity}
\rho = \frac{\sum_{(v_i,v_j)\notin E} \sum_{c_i\in C ,c_j\in C} max\Big(P(v_i|c_i) \times P(v_j|c_j),P(v_i|c_j) \times P(v_j|c_i)\Big)}{\sum_{(v_i,v_j)}\sum_{c_i\in C,c_j \in C} max\Big(P(v_i|c_i) \times P(v_j|c_j),P(v_i|c_j) \times P(v_j|c_i)\Big)}
\end{equation}

Figure \ref{fig:BlcokExample} depicts the main components in criterion \eqref{eq:main2}. After computing the internal association and external association degree of each node, final propagation probability of each node depends on the importance levels $p_1$ and $p_2$ obtained based on the following equations,
\begin{equation}
 \begin{cases}
\frac{p_1(c_i)}{p_2(c_i)} = \frac{\sum_{(v_i,v_j), v_i\in c_i,v_j\in c_i}(v_i,v_j)}{\sum_{(v_i,v_j), v_i\in c_i, v_j\notin c_i}(v_i,v_j)}\\
p_1(c_i) + p_2(c_i) = 1\\
\end{cases}
\label{eq:coef}
\end{equation}
\subsection{The description of the IEDC algorithm}
The details of the proposed approach is given  in Algorithm \eqref{alg:example}. 
The Algorithm \eqref{alg:example} initially depends on the structure of the network, number of iterations $MAXITER$ for updating the propagation probability. In this algorithm, first we initialize the propagation probability of each node to each community. The scalable edge clustering method is used  for getting the initial community structure of network \cite{tang_scalable_2012}. \emph{InteractionMatrix} and \emph{findFactors} are two functions for obtaining the interaction community matrix and the importance levels of internal and external association degree according to relations \eqref{eq:BlockMle} and \eqref{eq:coef}. Lines 7-22 explain the iterative framework for updating the propagation probability along with internal association degree and external association degree until to reach convergence. The \emph{findIA} function computes the internal association degree of each node to each community according to equation \eqref{eq:IA}. The \emph{findEA} function estimates the external association degree of each node based on relation \eqref{eq:Overlapping} and \emph{UpdatePropagation}, updates the propagation probability according to relation \eqref{eq:main2}. Finally, each node $v_i$ will be assigned to community $c_i$, if the propagation probability of node $v_i$ to community $c_i$ is greater than $threshold$ value.
\begin{algorithm}[t!]
	\caption{\small{Internal and External association to Detect Communities (IEDC)}}
	\label{alg:example}
	\begin{algorithmic}[1]
		\STATE {\bfseries Input:}  $G = (V,E), MAXITER$.
		\STATE {\bfseries Output:}  $F(c_i)$ Community memberships.
		\STATE {\bfseries Initialize:} Finding initial community propagation probability $P(v_i|c_i)$. 
		\STATE {\bfseries Compute:} $\beta(c_i,c_j) = InteractionMatrix(P,G)$
		\STATE {\bfseries Compute:} $p_1(c_i) , p_2(c_i) = findFactors(P,G)$
		\STATE {$Iter \gets 0$}
		\WHILE {$Iter < MAXITER$}
		\STATE {$Iter \gets Iter + 1$}
		\FOR {$i = 1$ \TO $N$}
		\FOR {$j=1$ \TO $k$}
		\STATE {$IA_{v_i}(c_j) = findIA(P,G)$ }
		\STATE {$EA_{v_i}(c_i) = findEA(P,G,\beta)$ }
		\STATE {\bfseries Update:}  $P^{update}(v_i|c_j)= UpdatePropagation(P,IA,EA,p_1,p_2)$.
		\ENDFOR
		\ENDFOR
		\ENDWHILE
		\FOR {$i = 1$ \TO $N$}
		\FOR {$j=1$ \TO $k$}
		\IF{$P(v_i|c_j)\geq threshold$}
		\STATE {\bfseries Add: }$F[c_j] \gets v_i $
		\ENDIF
		\ENDFOR
		\ENDFOR	
	\end{algorithmic}
\end{algorithm}
\subsection{The time complexity analysis of the proposed approach}
The \emph{IEDC} method is implemented on the three phases. In the first phase, the seed sets of communities are found based on the \emph{link clustering} approach \cite{tang_scalable_2012}. Since that the most of the real world networks have a sparse connectivity structures, the time complexity of the \emph{link clustering}  method is $O(m)$ where $m$ is the number of links in the network. The second phase is comprised the internal association degree computation for each node of the network based on Equation \ref{eq:IA}. The time complexity of this phase depends on the number of links and communities. It will take at most $O(n\times m\times k)$ based on fully connected network assumption where $n$ is the number of nodes and $k$ is the number of communities. Due to the sparsity of the real networks, the  computation time of the second phase is reduced to $O(m)$. Finally, the external association degree of nodes to each community is calculated based on Equation \ref{eq:Overlapping} in the third phase. The  Equation \ref{eq:Overlapping} could be computed in  $O(m)$ for the fixed value of $\beta$ parameter. In the worst scenario updating the  $\beta$  parameter  takes $O(n^2)$  due to the complexity of the denominator computation which it can be optimized based on  decomposing  Equation \ref{eq:BlockMle} into two factors as,
\begin{equation*}
\sum_{v_i, v_j}\big(P(v_i|c_i)  P(v_j|c_j)\big) = \sum_{(v_i, v_j)\in E}\big(P(v_i|c_i) P(v_j|c_j)\big) + \sum_{(v_i, v_j)\notin E}\big(P(v_i|c_i)  P(v_j|c_j)\big)
\end{equation*}
The first factor takes $O(E)$  and the second ones takes $O(n^2 - E)$ computation time. Finally, the upper bound of the time complexity would be $O(n^2)$.

\subsection{The sensitivity analysis of the IEDC algorithm}
\label{SecSensAnal}
The proposed algorithm consists of the computation of internal and external association degree for each node and these computations depend on some parameters that may affect the robustness and performance of the \emph{IEDC}. This approach recalls edge clustering technique to get the initializations of the communities memberships (also known as seed set configuration). In addition, the final probabilistic node assignment to a community requires a threshold specification. A series of experiments are conducted  to testify the robustness and sensitivity of the \emph{IEDC} to the initialization clustering step and the threshold specification in the final node membership computation. In the following,  first we assess the proposed approach based on different initialization clustering techniques, then investigate the performance of the other potential community detection methods along with the \emph{IEDC}  based on the same initialization strategy, and finally study the effect of  various  threshold selection strategies on the performance of the proposed approach.
\subsubsection{The initialization analysis}
We have designed a variety of experiments to assure the suitable selection of the edge clustering method \cite{tang_scalable_2012} to give good initialization seed set assignment for the community membership of the nodes. To this aim, first the \emph{IEDC} is tested through three well-known  clustering initialization methods, including  the spectral clustering method, \emph{Spectral} \cite{newman_modularity_2006}, the conductance based method, \emph{Conductance} \cite{gleich2012vertex}, and the equal weight for each community assignment to the node's approach, \emph{equal} \cite{airoldi_mixed_2008}, to analyze the sensitivity of the proposed method on this initialization stage. 
\begin{table}[t!]
	\centering
	\small
	\caption{The information about the networks  applied in the sensitivity analysis experiments }
	\label{tbl:senanal}
	\small
	\resizebox{0.63\textwidth}{!}{
		\begin{tabular}{lccc}
			\toprule[1.5pt]
			Network & Nodes.No & Links.No & Communities.No\\ 
			\midrule 
			\textit{Youtube} \cite{yang_defining_2015} & 1,134,890 & 2,987,624 & 8385 \\
			\textit{Orkut} \cite{yang_defining_2015} & 3,072,441 & 117,185,083 & 6,288,863 \\
			\textit{Livejournal} \cite{yang_defining_2015} & 3,997,962 & 34,681,189 & 287,512\\
			\textit{Polbooks} \cite{newman2006modularity} & 105  & 441 & 3  \\
			\textit{Polblogs} \cite{adamic2005political} & 1490 & 16726 & 2 \\
			\textit{CalTech} \cite{traud2012social} & 769 & 16656 & 9 \\
		    \textit{LFR} \cite{lancichinetti2008benchmark} & 1000 & 14782 & 15\\
			\bottomrule[1.5pt]    
		\end{tabular} 
	}
\end{table}

\begin{figure}[h!]
	\centering
	\includegraphics[width=1\linewidth]{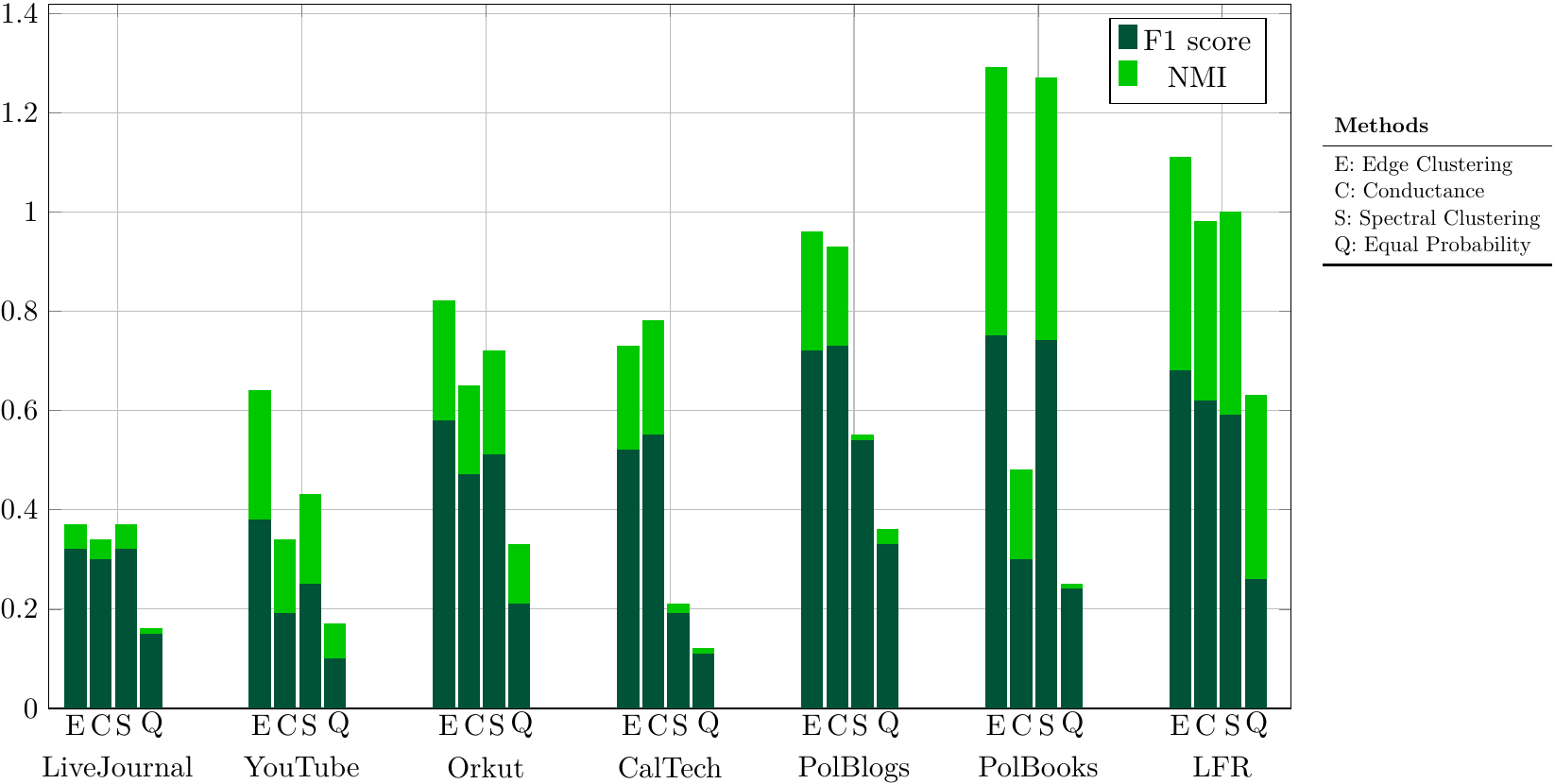}
	\caption{The effect of different clustering initialization methods on the proposed approach }
	\label{fig:sencluster}
\end{figure}

The real network datasets are considered ranging from small to large in Table \ref{tbl:senanal}, including the \emph{YouTube}, \emph{Orkut}, \emph{Livejournal}, \emph{PolBooks}, \emph{PolBlogs}, and \emph{CalTech}, and  an artificial simulated network based on \emph{LFR} method. It can be observed from Figure \ref{fig:sencluster} that the proposed method initialized based on $Edge$ clustering technique has better performance as compared with the other initialization methods. While in some datasets, the \emph{Spectral} and \emph{Conductance} have slightly better results, the $Edge$ clustering method benefits from linear time complexity as compared to these initialization procedures. This experiment concludes that the $Edge$ initialization method can be used as a suitable tool for extraction of the seed set for communities due to its better performance and less computational complexity than the others.\par
Moreover, we  investigate  the accuracy of the other community detection methods based on the same edge clustering initialization approach as well as the \emph{IEDC} method.
\begin{figure}[h!]
	\centering
	\includegraphics[width=1\linewidth]{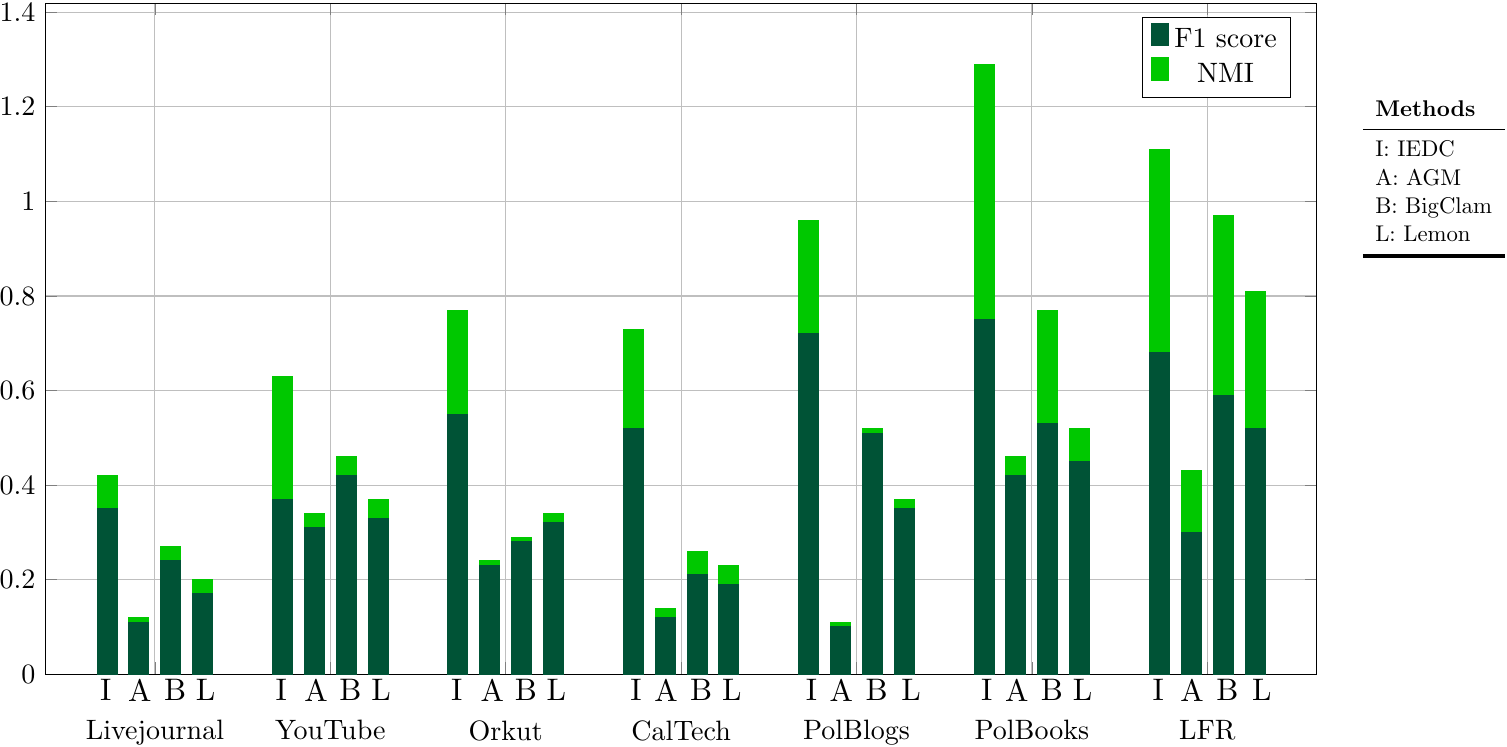}
	\caption{The comparison of initial based community detection approaches based on the same $Edge$ clustering initialization procedure}
	\label{fig:senclusothers}
\end{figure}
 The three well-known community detection methods  \emph{AGM}\cite{yang2012community}, \emph{BigClam}\cite{yang2013overlapping}, and \emph{LEMON}\cite{li_uncovering_2015} are selected among the others due to non-hierarchical type of the methods and dependency on the initial seed set of the communities.  
The obtained  experimental results in Figure \ref{fig:senclusothers} based on the dataset of Table \ref{tbl:senanal}  reveal that the \emph{IEDC} method outperforms the other techniques via the  \emph{NMI}  and \emph{F1 score} evaluation criteria along the same initialization approach. Moreover, the results of the other techniques have less accuracy than their default initialization procedures.  This result ensures us that the further investigation of the proposed method along the other community detection approaches via their default setting of the communities seed sets are fair enough as compared with our initialization approach.\par
\subsubsection{The threshold parameter}
The threshold selection is an important problem in supervised and unsupervised learning problems and is discussed slightly in some related community detection works \cite{hric_community_2014, ding_overlapping_2016, rees_overlapping_2012}. To select a threshold without prior knowledge of the network and running multiple experiments to fine-tune the threshold selection, we have selected three different automatic strategies for threshold determination including the  maximum value of  membership probability among  all the communities \emph{maximum}, assigning to the community with greater  membership probability than the average values for  all the nodes \emph{Average of All}, assigning to the community with greater  membership probability of a node  than the average values of its neighbors \emph{Average of Neighbors}.

\begin{figure}[h!]
	\centering
	\includegraphics[width=1\linewidth]{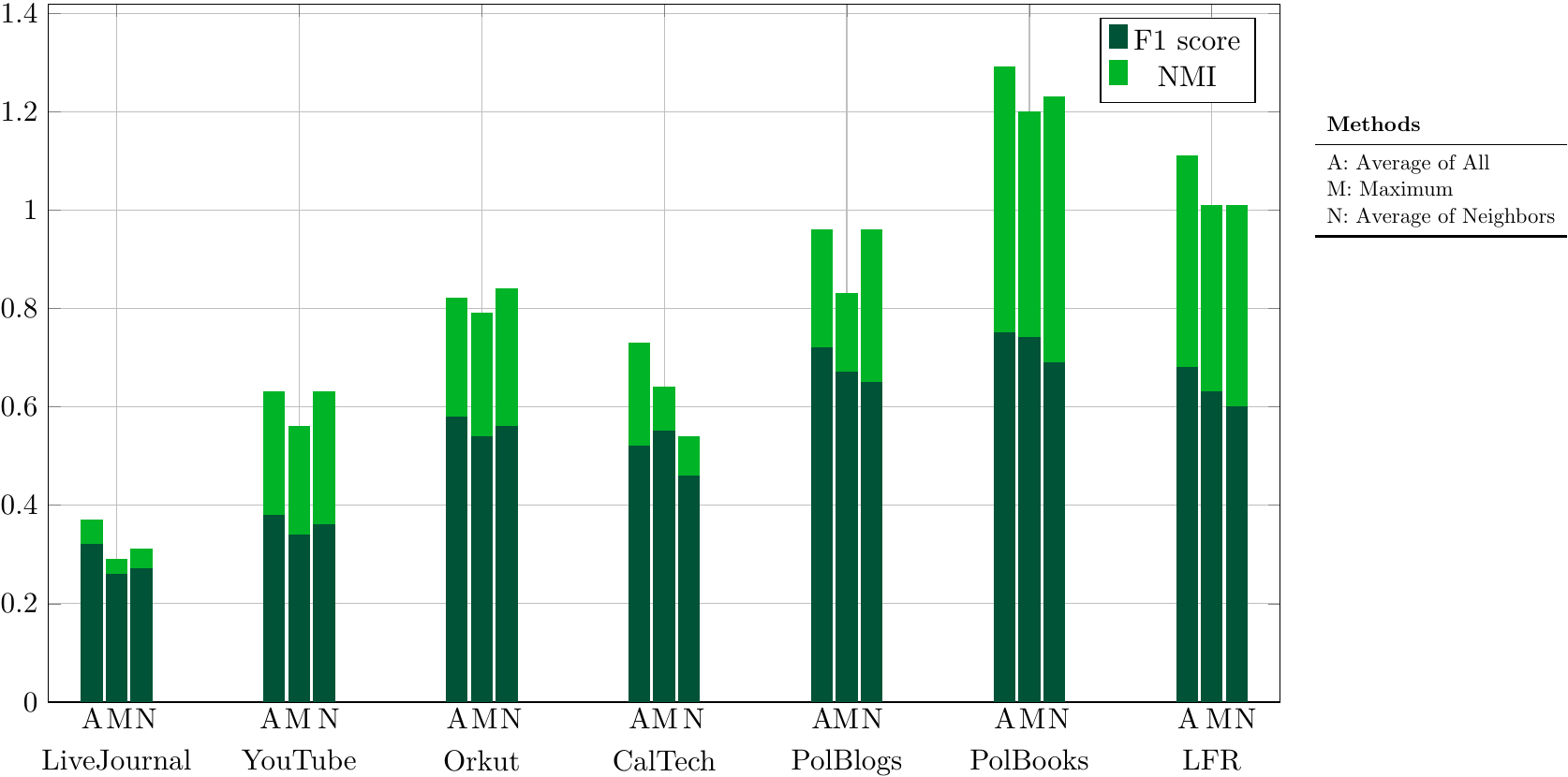}
	\caption{The investigation of different threshold selection strategies on the proposed method}
	\label{fig:SenThreshold}
\end{figure}

Here through the experiments on the datasets of Table \ref{tbl:senanal}, we  investigate the effects of  different threshold choices on the performance of \emph{IEDC} algorithm.  The results in Figure \ref{fig:SenThreshold} show that the proposed method is not highly sensitive to these threshold choices. Furthermore, the \emph{IEDC} had slightly better performance based on \emph{Average of All} threshold strategy than the others. Hence \emph{Average of All} is given as the default choice for threshold setting in our proposed configuration. 

\section{Experiments}
\label{Sec5}
We  have used a bunch of artificial generated networks and benchmark real datasets to analyze the IEDC algorithm.
At first, the proposed method is examined through extensive simulated artificial networks. Then, the accuracy of IEDC is compared with the earlier well-known algorithms based on real network datasets. 

\subsection{Evaluation criteria}
\label{EvalCriteria}
	The performance of the applied algorithms in our study are investigated based on two well-known evaluation criteria, the \emph{F1 score} and the Normalized Mutual Information (NMI), \cite{fortunato2010community}.  \emph{F1 score} measures the correctly classified members in each community based on the ground-truth information.  \emph{NMI}  criterion measures the similarity of the detected communities and the ground truth communities based on mutual information theoretic tools. There exist a variety of other evaluation criteria for community detection tasks including the maximization of the modularity function to detect the  non-overlapping community structures \cite{newman2006modularity} and modification of the modularity criteria to discover the overlapping community structures  \cite{li_detecting_2013}. One can see \cite{fortunato_community_2016} for the exact definitions of the aforementioned criteria. 
\subsection{Artificial networks generation } 
\label{NetworkGeneration}
To generate the artificial networks satisfying in a wide variety of situations,  two well-known approaches are exploited here, the Mixed Membership Stochastic Block-models approach, named as  \emph{MMSB} \cite{airoldi_mixed_2008} and the LFR method \cite{lancichinetti2008benchmark}. 
\begin{table}[H]
	\centering
	\small
	\caption{The details of artificial networks generation based on MMSB method}
	\label{tbl:MMSB}
	\resizebox{.95\textwidth}{!}{
		\begin{tabular}{lccccc}
			\toprule[1.5pt]
			Nodes.No & Links.No & Communities.No & Modularity & Type (dense/sparse) & Hyperparameter \\ 
			\midrule 
			100 & 1021 & 5 & 0.71 & Sparse & 0.003 \\ 
			100 & 1277 & 5 & 0.37 & Dense & 0.03  \\
			200 & 4635 & 11 & 0.26 & Dense & 0.05 \\
			200 & 2224 & 11 & 0.71 & Sparse & 0.002 \\
			500 & 6074 & 32 & 0.32 & Dense  & 0.02 \\
			500 & 4277 & 32 & 0.81 & Sparse & 0.001\\ 
			\bottomrule[1.5pt]    
		\end{tabular} 
	}
\end{table}


\begin{figure}[H]
	\centering
	\begin{minipage}{0.35 \textwidth}
		\centering
		\includegraphics[width=0.9\linewidth]{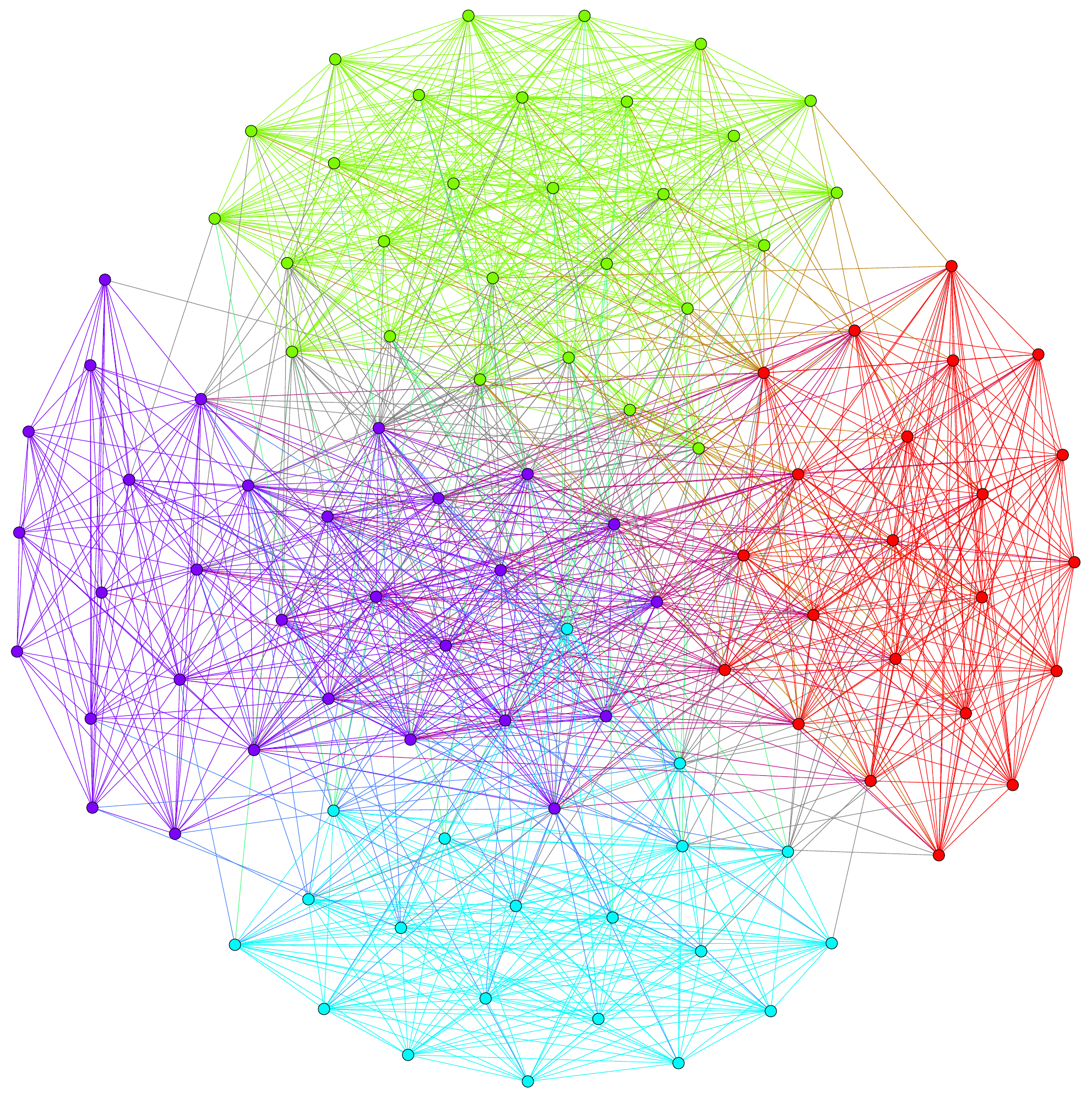}
		\subcaption{}
		\label{subfig:MMSB_Sparse}
	\end{minipage}%
	\begin{minipage}{0.35\textwidth}
		\centering
		\includegraphics[width=0.9\linewidth]{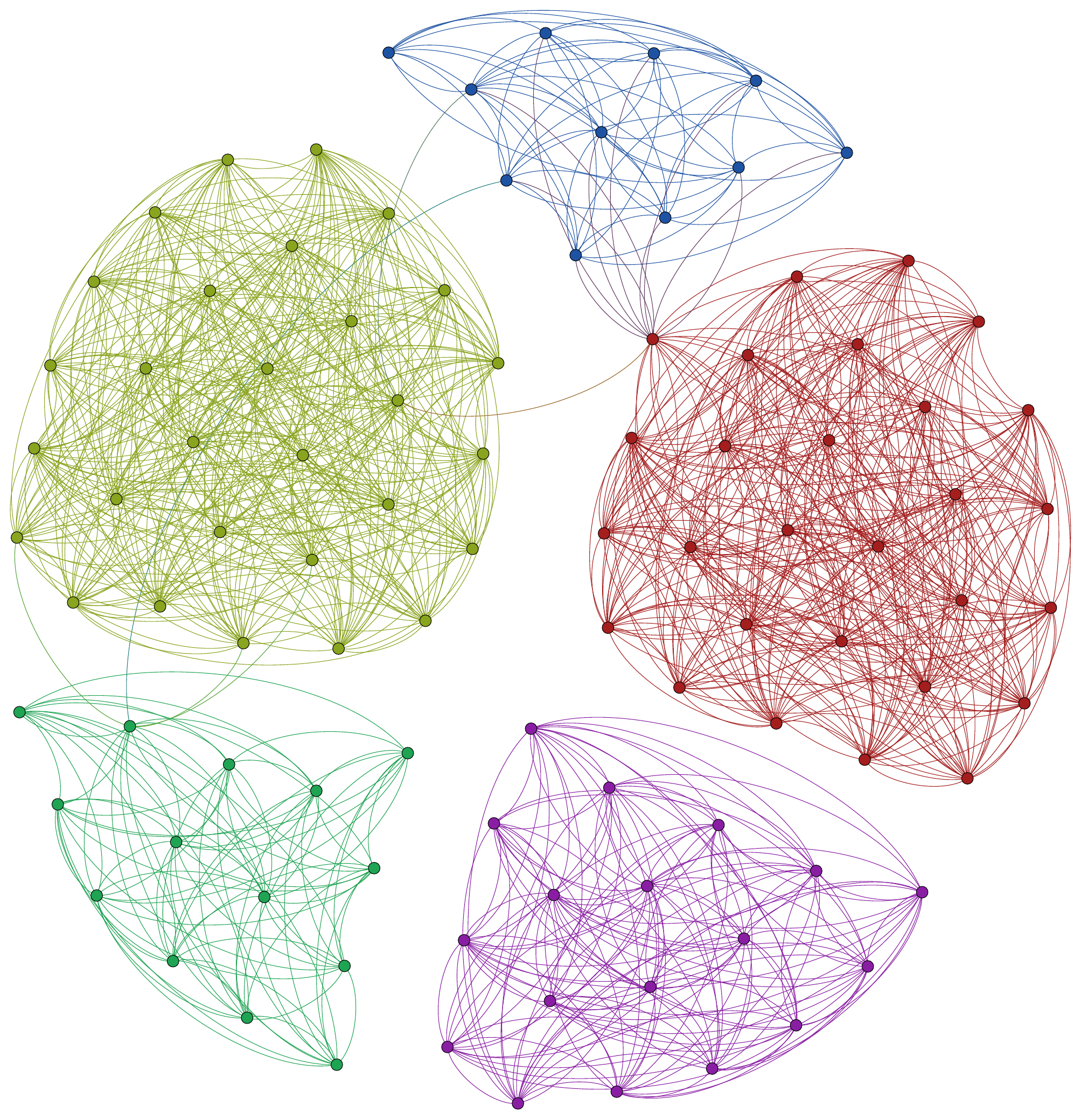}
		\subcaption{}
		\label{subfig:MMSB_Dense}
	\end{minipage}
	\\
	\begin{minipage}{0.35\textwidth}
		\centering
		\includegraphics[width=0.9\linewidth]{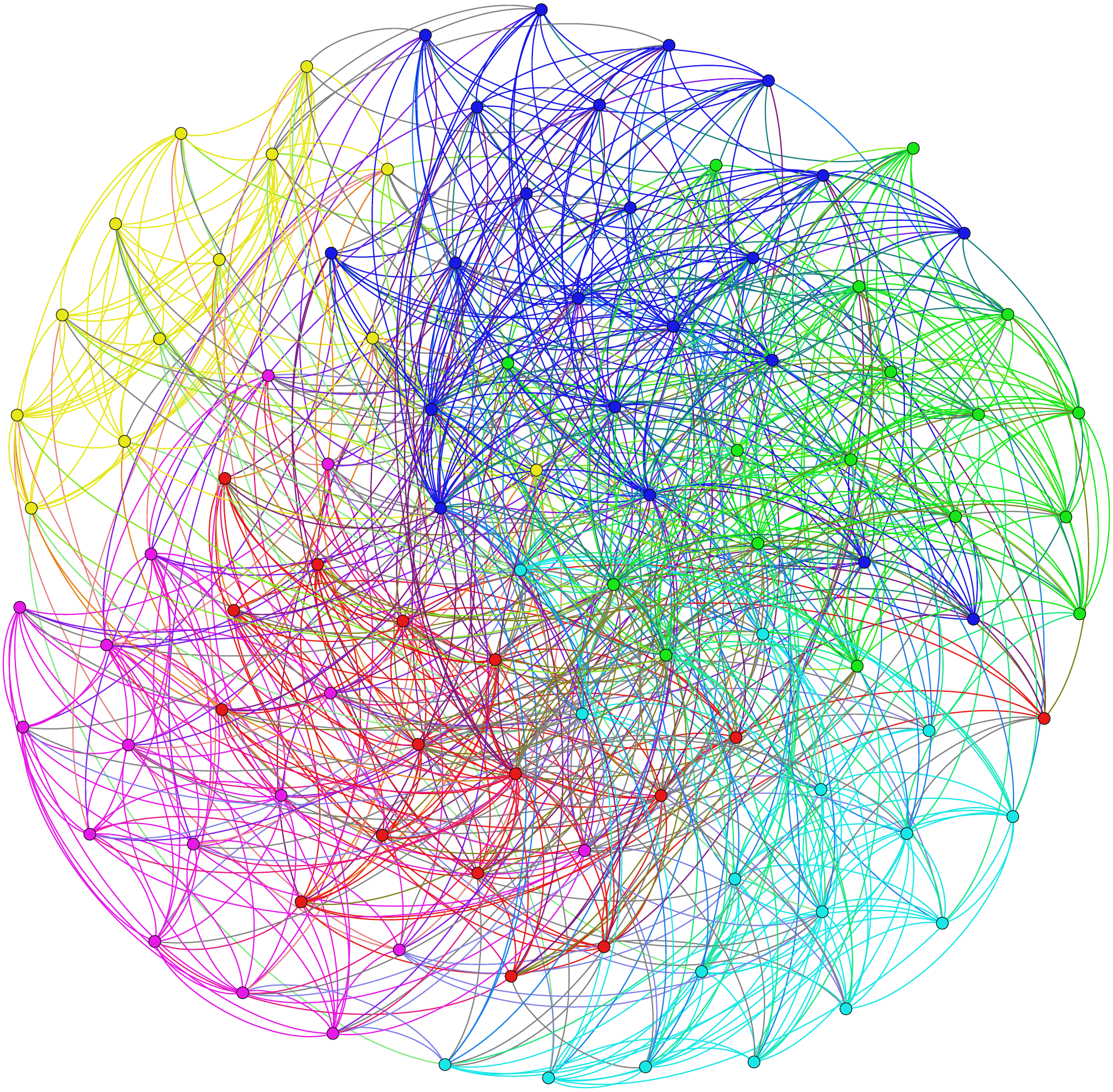}
		\subcaption{}
		\label{subfig:LFR_Dense}
	\end{minipage}%
	\begin{minipage}{0.35\textwidth}
		\centering
		\includegraphics[width=0.9\linewidth]{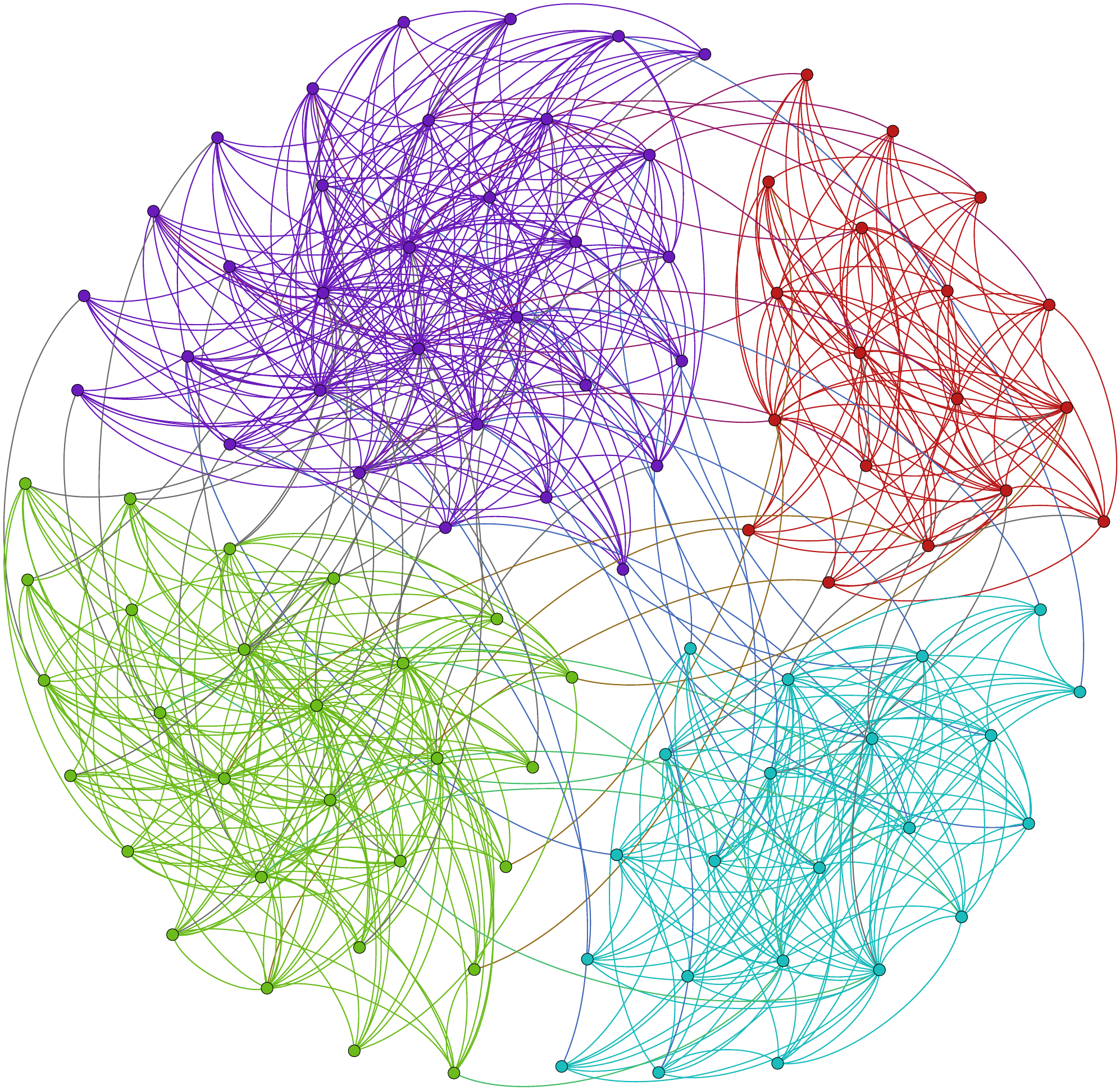}
		\subcaption{}
		\label{subfig:LFR_Sparse}
	\end{minipage}
	\caption{Artificial networks generation according to LFR and MMSB methods with 100 nodes. \ref{subfig:MMSB_Sparse} and  \ref{subfig:MMSB_Dense} show Dense and Sparse overlap communities with MMSB with $\alpha = 0.001$ and $\alpha = 0.07$.  \ref{subfig:LFR_Dense} and  \ref{subfig:LFR_Sparse}  represent Dense and Sparse overlap communities  with LFR according to Table \ref{LFR_parameter}. }
	\label{fig:LFRGenration}
\end{figure}

To study the performance of the IEDC algorithm versus the other  algorithms, the networks are artificially generated with two general scenarios,  sparse  and  dense overlap community structures  e.g. see Figure \ref{fig:LFRGenration}.\par
The MMSB procedure for network generation is built  upon in a probabilistic vein such that the link formation between two nodes $p$ and $q$ within a network, denoted by $Y(p, q)$,  are assumed to be distributed as, 
\begin{equation}
Y(p,q) \sim Bernoulli(Z_{p\rightarrow q}^T \beta Z_{p\leftarrow q})
\end{equation}
where  $\beta$ is a community interaction matrix  and $Z$ is a multinomial distribution as,
\begin{align*}
Z_{p\rightarrow q}&\sim Multinomial(\overrightarrow \pi_q)\\
\overrightarrow\pi_p &\sim Dirichlet(\overrightarrow \alpha) 
\end{align*}
The parameter $\alpha$ controls the overlapping behavior of communities. While the decreasing the value of $\alpha$ near to $0$ resulted in formation of networks with sparse overlapping behavior of  community structures, the increasing value of $\alpha$ near to $1$ tends to  formation of dense overlapping community structures within a network. The different behavior of $\alpha$ are shown in Figure \ref{fig:LFRGenration}. Due to the complexity of different behavior of $\alpha$,  the \textit{modularity} metric  applies to categorize  the levels of overlapping behavior among the communities \cite{newman2006modularity}. 
Based on our experimental studies,  the modularity values greater than $0.5$  leads to formation of sparse overlapping communities and  the modularity values less than $0.4$ resulted in dense overlap community structures.  Table \ref{tbl:MMSB} shows the characteristics of generated networks  according to the MMSB simulation approach. \par
\begin{table}[t!]
	\caption{The characterization of LFR procedure for network generation}
	\small
	\resizebox{0.95\textwidth}{!}{
		\begin{tabular}{lcccccc}
			
			\toprule[1.5pt]
			Nodes.No & Links.No&  Communities.No & Modularity & Type  & Mixing & Overlapping of nodes\\ 
			\midrule 
			100 & 963 & 6 & 0.336 & Dense & 0.3 & 20 \\ 
			100 & 727 & 4 &  0.605 & Sparse &  0.1 & 10  \\
			200 & 1924 & 10 & 0.277 &  Dense & 0.5 & 40 \\
			200 & 1503 & 11 &  0.6 & Sparse & 0.2 & 20 \\
			500 & 6123 & 16 &  0.306 & Dense  & 0.5 & 100 \\
			500 & 6224 & 16 &  0.61 & Sparse & 0.2 & 50\\ 
			\bottomrule[1.5pt]		
		\end{tabular} 
	}
	\label{LFR_parameter}
\end{table}
The LFR procedure is a standard framework  for generating  benchmark artificial networks to test the community detection algorithms \cite{lancichinetti2008benchmark}.  Similar to the MMSB approach,  the LFR generated networks are divided into two types of  sparse and  dense overlapping communities according to the  modularity metric. 
The details of network generation through the LFR approach is given  in Table \eqref{LFR_parameter}. The two typical generated networks with the LFR  is shown in Figure \ref{fig:LFRGenration}. 

\subsection{Experimental results based on artificial networks}
Initially, the conductance measure is used to justify  the  performance of  the proposed algorithm  based on the generated networks with the scenarios described in Subsection \ref{NetworkGeneration}. Then the \emph{IEDC} algorithm  is evaluated through the \emph{MMSB}  generated networks to check the feasibility of the approach. Finally, we represent the evaluation results of the proposed approach based on the \emph{LFR} to generate the artificial networks  for community detection tasks  through the state-of-the-arts community detection algorithms.\par
\begin{figure}[H]
	\centering
	\begin{minipage}{0.45 \textwidth}
		\centering
		\includegraphics[width=0.9\linewidth]{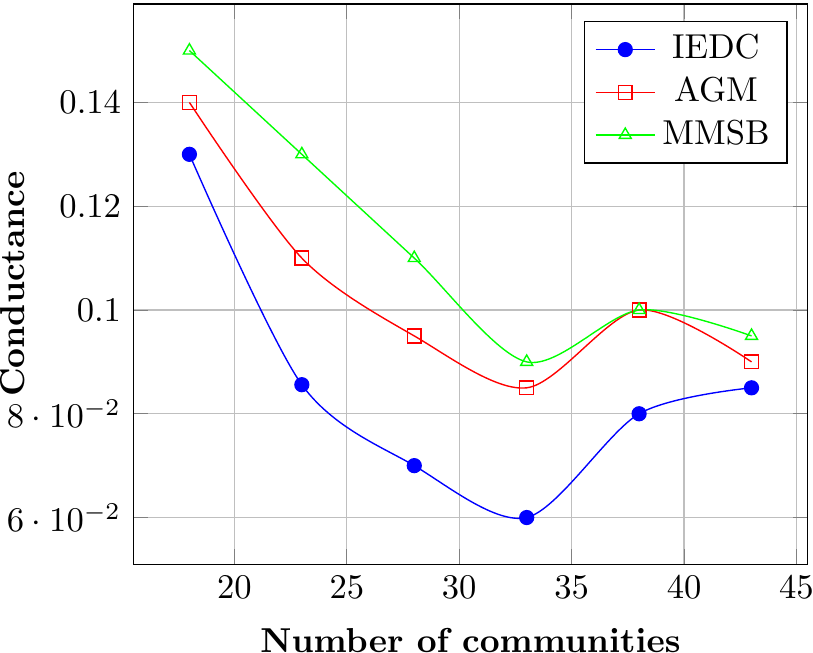}
		\subcaption{\scriptsize{Dense MMSB}}
		\label{subfig:MMSB_500_Sparse}
	\end{minipage}%
	\begin{minipage}{0.45\textwidth}
		\includegraphics[width=0.9\linewidth]{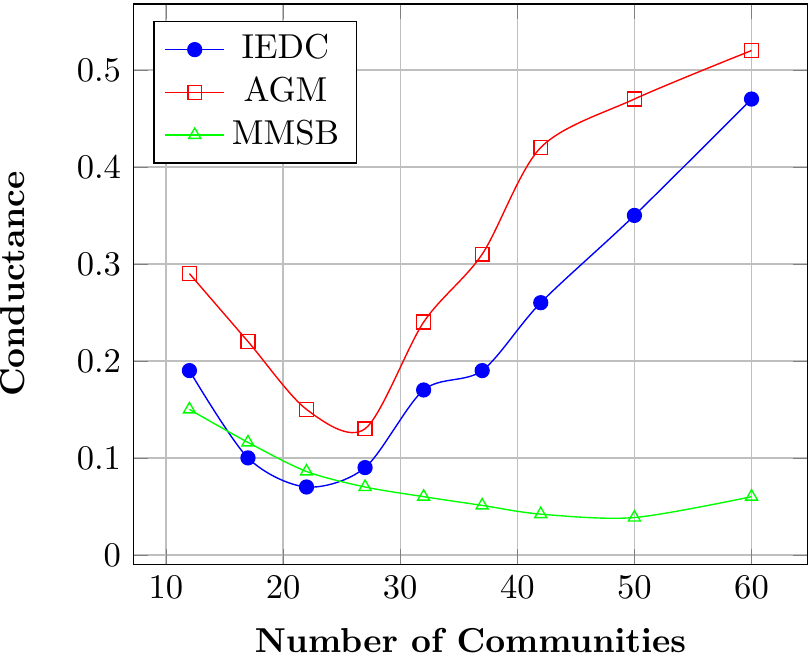}
		\subcaption{\scriptsize{Sparse MMSB}}
		\label{subfig:MMSB_500_Dense}
	\end{minipage}
	\\
	\begin{minipage}{.45\textwidth}
		\centering
		\includegraphics[width=0.9\linewidth]{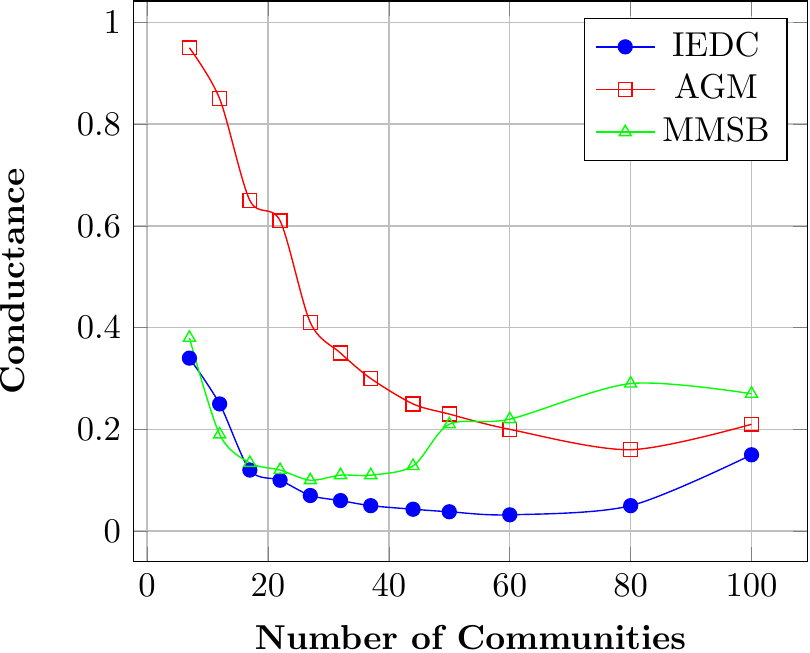}
		\subcaption{\scriptsize{Dense LFR}}
		\label{subfig:LFR_500_Sparse}
	\end{minipage}%
	\begin{minipage}{.45\textwidth}
		\centering
		\includegraphics[width=0.9\linewidth]{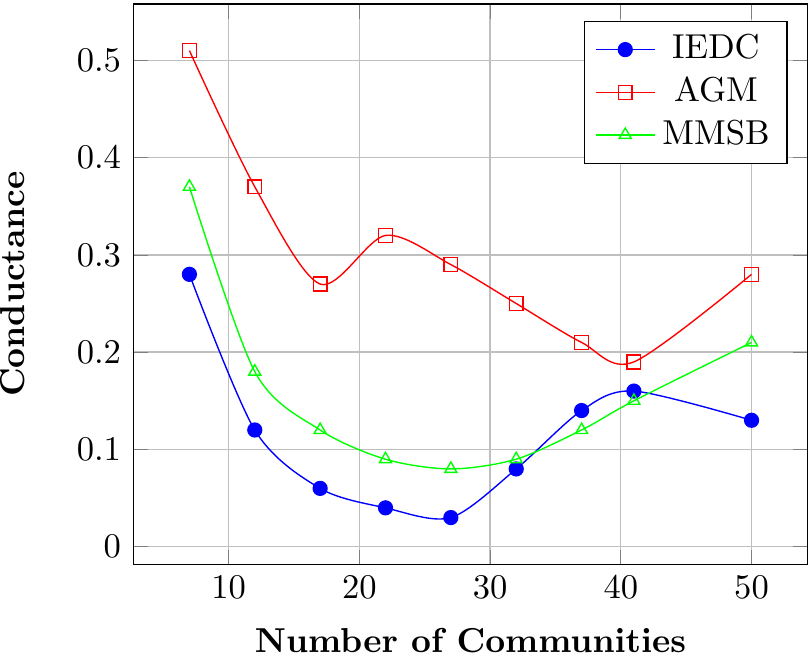}
		\subcaption{\scriptsize{Sparse LFR}}
		\label{subfig:LFR_500_Dense}
	\end{minipage}
	\caption{The Conductance measure based on the number of communities in the simulated networks. Sub-figures \ref{subfig:MMSB_500_Sparse} and  \ref{subfig:MMSB_500_Dense} show the Conductance level on the MMSB network generation with 500 nodes with dense   and  sparse overlapping communities assumption.  Sub-figures \ref{subfig:LFR_500_Sparse} and \ref{subfig:LFR_500_Dense} present the conductance level of the LFR networks with dense and  sparse overlapping assumption.}
	\label{fig:LFRMMSB-Conductance}
\end{figure}

 
In addition, we evaluate the IEDC algorithm with MMSB and AGM through the conductance measure depicted in  Figures \ref{fig:LFRMMSB-Conductance}. 
The conductance measure  is defined for the $i$-th community in a network as the ratio between  the cut size of that community and  the least number of links incident on either set $c_i$ or $V \backslash c_i$ \cite{gleich2012vertex}, 
\begin{equation}
\label{eq:conductance}
Cond(c_i) =  \dfrac{\cut(c_i)}{ \min \bigg(  \links(c_i, V), \links(V \backslash c_i, V) \bigg)}
\end{equation}	
where the $\links(c_i, V)$ is defined  the sum of edge weights between node sets $c_i$ and $V$.
The probability of leaving the community by a one-step walk beginning from the smaller set between $c_i$ and 
$V \backslash c_i$ can be interpreted as the  conductance of that community. Intuitively, the conductance measures the proportion of external edges of a  community versus the total degree of that community \cite{fortunato_community_2016}.
The successful minimization of the conductance would be effectively resulted in the two requirement of community structure, including  well-separated from the rest of the network (small numerator) and large internal density (large denominator).  
On the one hand, this measure is relatively insensitive to the size of the communities due to the proportional dependency of  both the numerator and the denominator to the number of nodes in a given community \cite{leskovec_community_2009}. On the other hand, by increasing the number of communities  it would be resulted in the community structures with fewer number of edges in each community and vice versa \cite{gleich2012vertex}. Therefore, we have used a modified version of the conductance measure for overall comparison among different community structures as the average of conductance of the communities in a partition of the network. The detailed study in \cite{leskovec_community_2009} have shown that the minimum conductance size versus the number of nodes in a community have a characteristic shape for a partition, the decreasing behavior  until $k \approx 100 $ nodes, and then increasing monotonically for larger subgraphs. Based on the correspondence between the size and the number of communities, the average of conductance measure for a partition is plotted against the number of communities in hopes of appearance a similar characteristic curve.  Intuitively, the conductance level of  networks tends to have a general  curve where initially it has a decreasing behavior versus to the increasing values of number of communities, and after some stage, the conductance level gradually raised with the increasing number of communities. 
We exploit the conductance measure versus the number of communities to evaluate the proposed method. 
The results in Figures \ref{subfig:MMSB_500_Sparse} and \ref{subfig:MMSB_500_Dense}, show us the power of the \emph{IEDC} method based on the conductance evaluation criterion for the MMSB network generation procedure as compared with the other methods. On the MMSB  generated networks, the \emph{IEDC} behavior is very similar to the \emph{AGM} method.  The Figures \ref{subfig:LFR_500_Sparse} and \ref{subfig:LFR_500_Dense}, present the results of the conductance measure for the LFR network generation procedure on the three studied methods, where the \emph{IEDC} seems to behave more normally as compared to the \emph{MMSB} and \emph{AGM} due to the well-shape of the IEDC curve in these cases to the recommended conductance shape \cite{gleich2012vertex}.\par
\begin{figure}[t!]
	\centering
	\includegraphics[scale=.8]{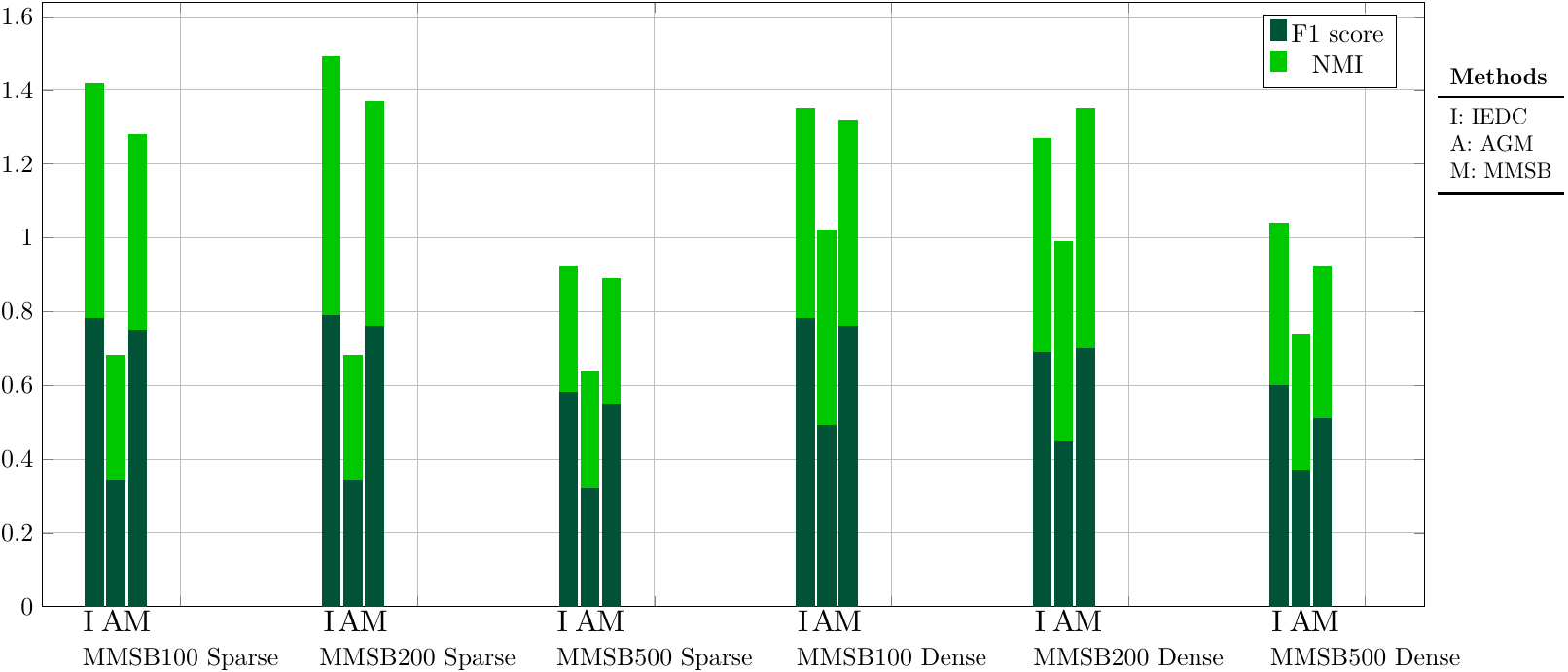}
	\caption{\label{fig:f1NMI-artificial-MMSB}The \emph{F1 score} and  \emph{NMI}  evaluation criteria based on running the algorithms on the MMSB simulated netwroks.}
\end{figure}
We perform the initial feasibility and justification of  the proposed approach  through \emph{MMSB} generated networks based on the   \emph{NMI}   and \emph{F1 score} as standard quantitative evaluation criteria.
Figure \ref{fig:f1NMI-artificial-MMSB} shows the   \emph{NMI}   and \emph{F1 score} results for simulated networks with  \emph{MMSB}  generation procedures based on  Table \ref{tbl:MMSB}. We observe that the proposed method performs superior or equally for most of the cases in sparse and overlap generated networks in the \emph{MMSB} generation scenario's. \par
\begin{figure}[H]
\centering
\includegraphics[scale=.75]{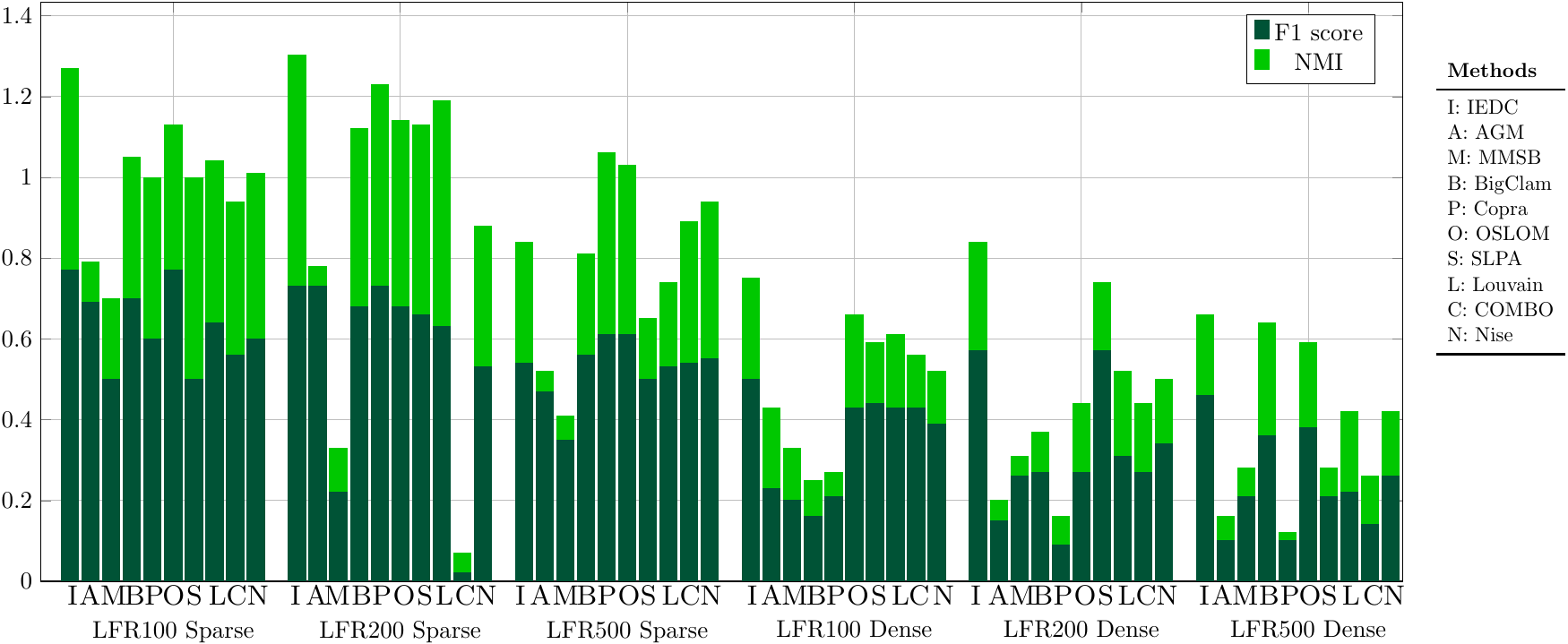}
\caption{The  \emph{NMI}  and \emph{F1 score} evaluation results on  the LFR simulated networks based on the well-known community detection algorithms in Table \ref{tbl:GeneralMethod}.}
\label{fig:f1NMI-artificial-LFR}
\end{figure}

\begin{table}[t!]
	\centering
	\small
	\caption{Overview of the state-of-the-art algorithm}
	\resizebox{.9\textwidth}{!}{
		\begin{tabular}{llc}
			\toprule[1.5pt]
			Methods & Description & Reference\\ 
			\midrule 
			MMSB & Find Overlapping community&\cite{airoldi_mixed_2008}\\
			AGM & Find Overlapping community & \cite{yang2012community}\\
			BigClam & Find Overlapping community & \cite{yang2013overlapping}\\
			SLPA & Find Overlapping community & \cite{xie2012towards}\\
			OSLOM & Find Overlapping Community, Hierarchies and community dynamics & \cite{lancichinetti2011finding} \\
			COPRA & Find Overlapping community by label propagation & \cite{gregory_finding_2010} \\ 
			Louvain & Heuristic method for finding non-overlapping community & \cite{blondel2008fast}\\
			COMBO & A general approach to detect the overlapping and non-overlapping communities & \cite{sobolevsky_general_2014-3}\\
			NISE & Find Overlapping community based on the seed set expansion& \cite{whang_overlapping_2016}\\    
			\bottomrule[1.5pt]
		\end{tabular} 
	}
	\label{tbl:GeneralMethod}
\end{table}
The \emph{LFR} is known as the standard simulation procedure to test the community detection algorithms. We employ  several state-of-the-art community detection algorithms to examine the \emph{IEDC} approach including the scalable version of  the \emph{AGM}, \emph{BigClam} \cite{yang2012community}, greedy hierarchical method based on modularity maximization,  \emph{Louvain} \cite{blondel2008fast},  label propagation based approach,  \emph{COPRA} \cite{gregory_finding_2010},  Speaker-listener Label Propagation Algorithm,  \emph{SLPA} \cite{xie2012towards} and  Order statistics local optimization method, \emph{Oslom} \cite{lancichinetti2011finding}, optimize multiple objective functions based on a general search strategy, \emph{COMBO} \cite{sobolevsky_general_2014-3}, and  Neighborhood-Inflated Seed Expansion technique, \emph{NISE} \cite{whang_overlapping_2016} summarized in Table \ref{tbl:GeneralMethod}.  The results for \emph{LFR} generated networks are represented in Figure \ref{fig:f1NMI-artificial-LFR} based on the characteristics in Table \ref{LFR_parameter}. The attained \emph{NMI}   and \emph{F1 score} results shows that the \emph{IEDC} outperforms the state-of-the-art community detection algorithms on Table \ref{tbl:GeneralMethod} for the whole simulation scenarios except the sparse generated network with $500$ nodes. These results have shown that the strength of the  proposed approach to detect the community structures based on the \emph{LFR} benchmark community detection framework. 

\subsection{Experiments on real network data-sets}
The proposed method is compared with the state-of-the-art community detection algorithms through a variety of benchmark real networks. We use two evaluation criteria  \emph{NMI}  and  \emph{F1 score} to examine the proposed approach on the real networks.
To more clarify, we categorize the networks under study to the  non-overlapping and overlapping ones based on the ground truth information about the community structures and report the results separately in the following subsections.
\begin{table}[t!]
	\centering
	\small
	\caption{The networks with non-overlapping community structures }
	\label{tbl:realworld}
	\small
	\resizebox{0.63\textwidth}{!}{
		\begin{tabular}{lccc}
			\toprule[1.5pt]
			Network & Nodes.No & Links.No & Communities.No\\ 
			\midrule 
			\textit{Football} \cite{girvan2002community}& 115 & 616 & 12 \\ 
			\textit{Polbooks} \cite{newman2006modularity} & 105  & 441 & 3  \\
			\textit{Polblogs} \cite{adamic2005political} & 1490 & 16726 & 2 \\
			\textit{CalTech} \cite{traud2012social} & 769 & 16656 & 9 \\
			\textit{Rice} \cite{traud2012social} & 4087 & 184828 & 10 \\
			\bottomrule[1.5pt]    
		\end{tabular} 
	}
\end{table}
\subsubsection{Networks with non-overlapping community structures} 
\label{SecRealNonverRes}
\begin{figure}[h!]
\centering
\includegraphics[scale=.8]{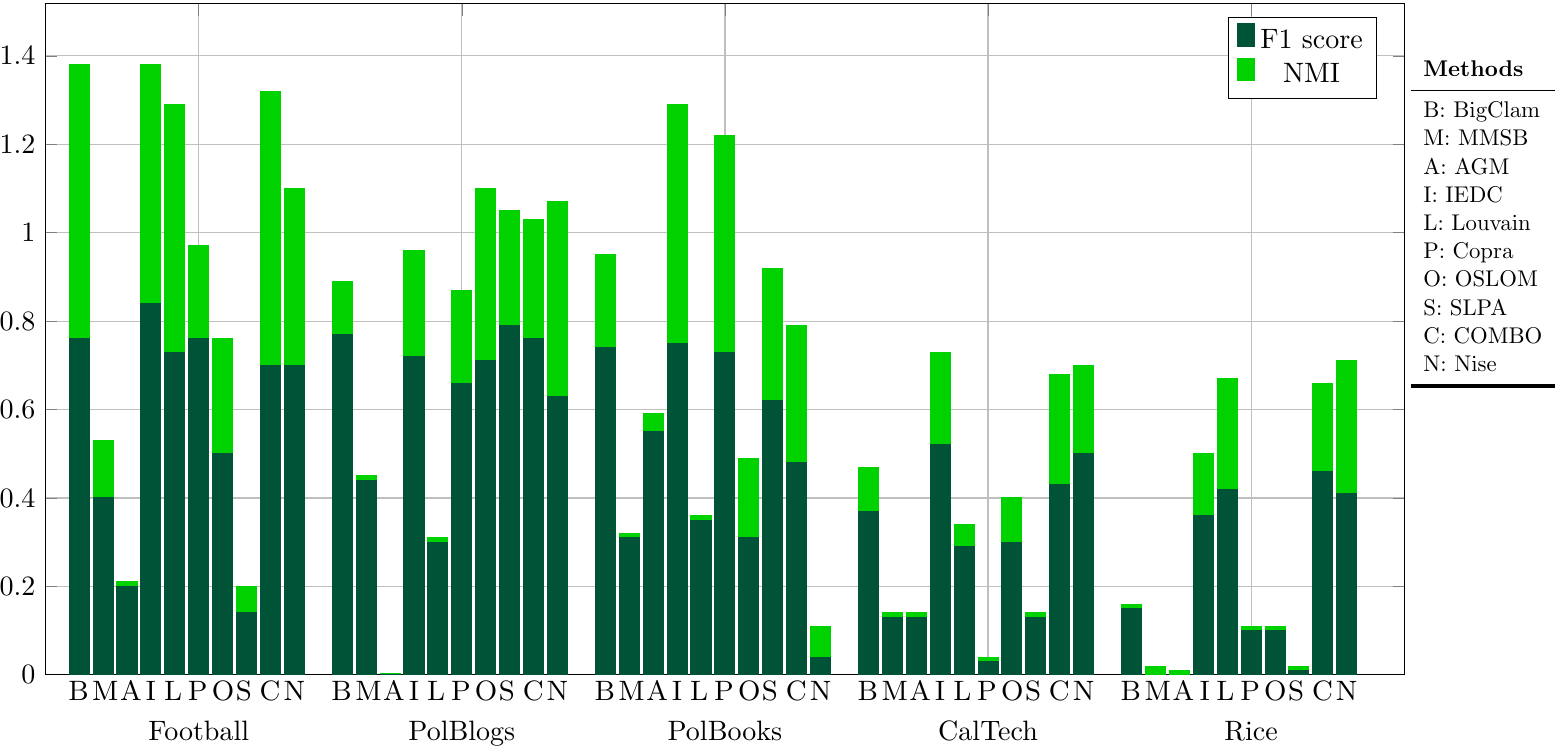}
\caption{The results of \emph{F1 score} and \emph{NMI} criteria  on networks in Table \ref{tbl:realworld} with non-overlapping community structures.}
\label{fig:EvaluationNonOverlap}
\end{figure}

Multiple real network datasets in Table \ref{tbl:realworld} with non-overlapping community structures  are used  to evaluate the proposed method. The experimental results are shown in Figure \ref{fig:EvaluationNonOverlap} based on two well-known evaluation criteria.\par
According to Figure \ref{fig:EvaluationNonOverlap},  \emph{IEDC} performs slightly better or equally  than the other algorithms based on \emph{F1 score} and  \emph{NMI}  on the \emph{Polbooks}, \emph{CalTech} and \emph{Football}. In addition,  the proposed method shows slightly weaker results on \emph{Rice} and {Polblogs} based on \emph{F1 score} and weak performance via NMI. The attained results show that the \emph{IEDC} approach is able to detect non-overlapping community structures with suitable accuracy as compared with the other well-known algorithms based on $5$ different network datasets.

\subsubsection{Networks with overlapping community structures} 
We describe the characteristics of $6$ real network datasets  in Table \ref{tbl:HugeNetworks} with overlapping community structures. The community detection algorithms in Table \ref{tbl:GeneralMethod} are used to examine the proposed approach with the exclusion of  \emph{AGM} and \emph{MMSB} algorithms due to the non-scalable and heavy time-consuming of these procedures for the large networks.  The experimental results on overlapping network datasets are represented  in Figure \ref{fig:EvaluationNonOverlap} based on    \emph{NMI}   and \emph{F1 score} evaluation criteria. 
\begin{table}[t!]
	\centering
	\small
	\caption{Characteristics of real world networks with overlapping community structures.}
	\resizebox{0.7\textwidth}{!}{
		\begin{tabular}{lcccc}
			\toprule[1.5pt]
			Network & Node.No & Links.No & Communities.No &  Type\\ 
			\midrule 
			\textit{BlogCatalog} & 10312 & 667,966 & 39 & \textit{Large}\\ 
			\textit{Facebook} \cite{leskovec2012learning} & 4089 & 170,174 & 193 & \textit{Large} \\
			\textit{DBLP} \cite{yang_defining_2015} & 317080 & 1,049,866 & 13477 & \textit{Huge}\\
			\textit{	Youtube} \cite{yang_defining_2015} & 1,134,890 & 2,987,624 & 8385 & \textit{Huge} \\
			\textit{Orkut} \cite{yang_defining_2015} & 3,072,441 & 117,185,083 & 6,288,863 & \textit{Huge} \\
			\textit{Livejournal} \cite{yang_defining_2015} & 3,997,962 & 34,681,189 & 287,512 & \textit{Huge}\\
			\bottomrule[1.5pt]
		\end{tabular} 
	}
	\label{tbl:HugeNetworks}
\end{table}
Due to the facing of the huge networks with overlapping community structures and non-scalability  running property of  the baseline algorithms on these large networks  in Table \ref{tbl:HugeNetworks} (except the \emph{BigClam} method), we have exploited a community-based sampling strategy to obtain overlapping subgraph communities in each networks \cite{yang2013overlapping}. To sample a subnetwork with overlapping community structures, a random node $u$ in a  network $G$  is selected subject to the constraint that it belongs to at least two communities. We then take the subnetwork of $G$ comprising all the nodes that share at least one community membership with the node $u$. We created $100$ subnetworks for the huge specified networks and reported the average result based on these subnetworks. 
\begin{figure}[H]
\centering
\includegraphics[scale=.8]{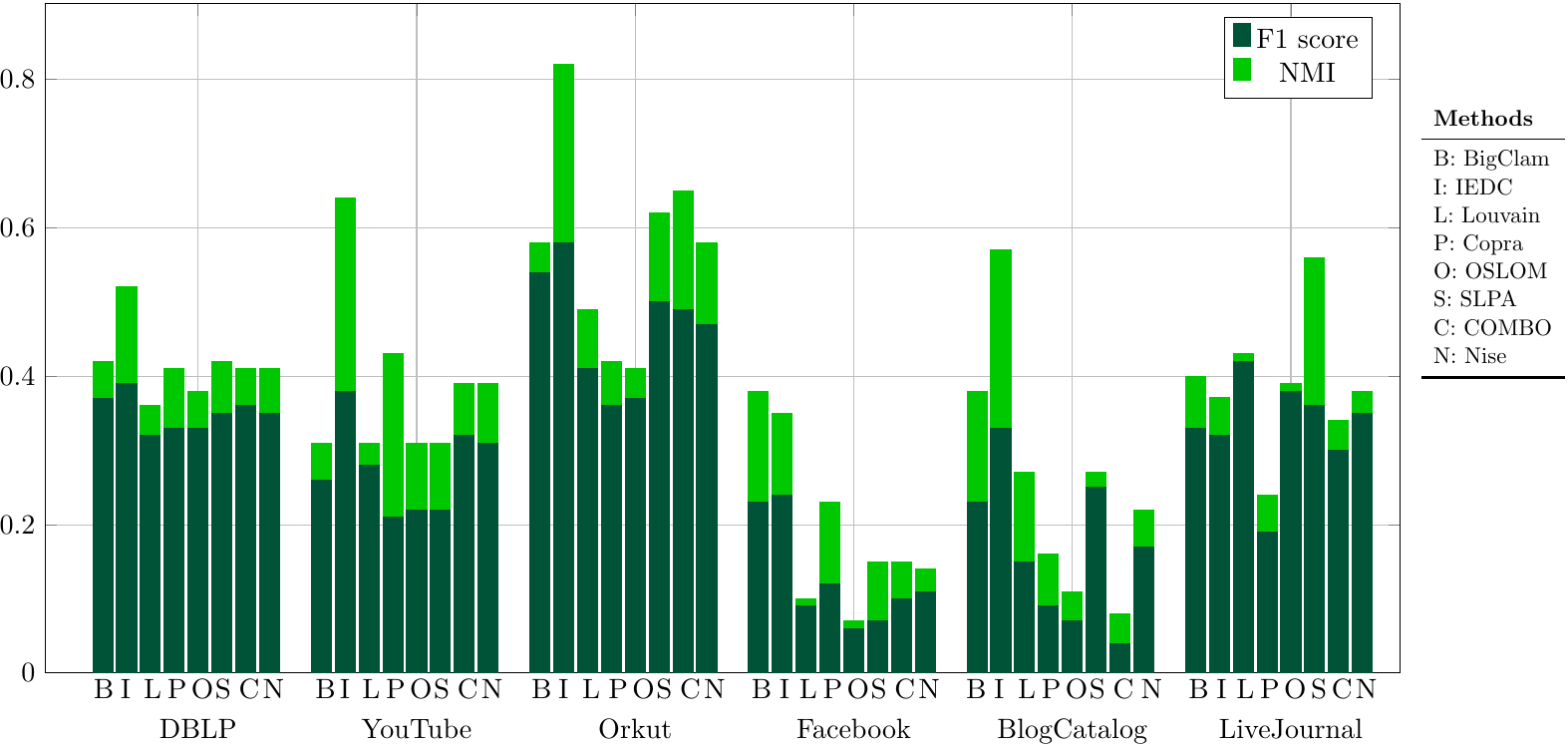}
\caption{The results of \emph{F1 score} and \emph{NMI} criteria  on the real world networks in Table \ref{tbl:HugeNetworks} with overlapping community structures.}
\label{fig:RealOverlapping}
\end{figure}
The Figure \ref{fig:RealOverlapping} depicts the   \emph{NMI}   and \emph{F1 score} results for the networks described in Table \ref{tbl:HugeNetworks}.
According to Figure \ref{fig:RealOverlapping},  \emph{IEDC} performs better than the other algorithms based on \emph{F1 score} on \emph{BlogCatalog}, \emph{DBLP}, \emph{YouTube} and \emph{Orkut} and slightly weak results  on the \emph{Livejournal}. 
The results based on   \emph{NMI}   metric reveals  that the \emph{IEDC} achieves  better results than the others on the \emph{BlogCatalog}, \emph{DBLP}, \emph{YouTube}, and \emph{Orkut}, also weaker results than the others on the  \emph{Livejournal} and \emph{Facebook}. On \emph{Orkut} network, \emph{IEDC} attained  two times better result versus the top result of the other methods. 
\section{Conclusion and future works}
\label{Sec6}
In this paper we developed a novel generalized community detection approach to discover the communities with overlapping and non-overlapping  connectivity structures. We have designed our \emph{IEDC} algorithm based on a primary node based criterion via the internal and external association degree computation for each node to detect communities. We tested the proposed method on extensive simulation experiments with two artificial generative procedure, the \emph{LFR} and the \emph{MMSB} procedures. The attained results on extensive simulated networks through two generative scenarios via the NMI, \emph{F1 score} and conductance measure as the evaluation criteria showed us the superior performance and feasibility of our proposed method as compared with the well-known algorithms.  We have used a bunch of real network data-sets with different overlapping community structures and a variety of small to huge benchmark networks. The experimental results verified the benefits of the \emph{IEDC} approach and its strengths for general community detection versus to the earlier state-of-the-art network communities algorithms.\\
There exist several directions for future works on this field including,
\begin{itemize}
\item Perform theoretical analysis of the proposed approach via the probabilistic method.
\item Study the other properties of the networks under study such as core-periphery based on the \emph{IEDC}.
\item Use  the profile information or attributes of each node along with the structure of network to detect generalized community structures.
\end{itemize}
\section*{References}

\end{document}